\documentclass[prx,twocolumn,amsmath,amssymb,superscriptaddress,longbibliography]{revtex4-2}

\usepackage{color,times}
\usepackage{dcolumn}
\usepackage{multirow}
\usepackage{amssymb,amsfonts,amsmath,graphicx}
\usepackage{epsfig}
\usepackage{xcolor}
\usepackage{SIunits}
\usepackage{bm}
\usepackage{braket}
\usepackage{dsfont}
\usepackage[utf8]{inputenc}
\usepackage{epstopdf}
\usepackage[english]{babel}
\definecolor{darkblue}{rgb}{0, 0, 0.8}
\usepackage[colorlinks=true, breaklinks=true, linkcolor=darkblue, citecolor=darkblue, urlcolor=darkblue]{hyperref}

\graphicspath{{..//Figures//}}

\definecolor{darkgreen}{rgb}{0, 0.5, 0}

\definecolor{tlc}{rgb}{0.5, 0.1, 0.5}

\definecolor{hred}{HTML}{E43E4C}
\usepackage[normalem]{ulem}
\usepackage{chngcntr}
\counterwithout{equation}{section}
\addtocounter{equation}{0}

\usepackage[normalem]{ulem}

\usepackage{chngcntr}
\counterwithout{equation}{section}
\addtocounter{equation}{0}

\begin{document}

\title{Spectroscopy of elementary excitations from quench dynamics in a dipolar XY Rydberg simulator}

\author{Cheng~Chen$^*$}
\affiliation{Universit\'e Paris-Saclay, Institut d'Optique Graduate School,\\
		CNRS, Laboratoire Charles Fabry, 91127 Palaiseau Cedex, France}
		
\author{Gabriel~Emperauger$^*$}
\affiliation{Universit\'e Paris-Saclay, Institut d'Optique Graduate School,\\
		CNRS, Laboratoire Charles Fabry, 91127 Palaiseau Cedex, France}		
		
\author{Guillaume~Bornet$^*$}
\affiliation{Universit\'e Paris-Saclay, Institut d'Optique Graduate School,\\
		CNRS, Laboratoire Charles Fabry, 91127 Palaiseau Cedex, France}
		
\author{Filippo~Caleca$^*$}
\affiliation{Univ Lyon, Ens de Lyon, CNRS, Laboratoire de Physique, F-69342 Lyon, France}

\author{Bastien~G\'ely}
\affiliation{Universit\'e Paris-Saclay, Institut d'Optique Graduate School,\\
		CNRS, Laboratoire Charles Fabry, 91127 Palaiseau Cedex, France}

\author{Marcus~Bintz}
\affiliation{Department of Physics, Harvard University, Cambridge, Massachusetts 02138 USA}

\author{Shubhayu~Chatterjee}
\affiliation{Department of Physics, Carnegie Mellon University, Pittsburgh, PA 15213, USA}

\author{Vincent~Liu}
\affiliation{Department of Physics, Harvard University, Cambridge, Massachusetts 02138 USA}

\author{Daniel~Barredo}
\affiliation{Universit\'e Paris-Saclay, Institut d'Optique Graduate School,\\
		CNRS, Laboratoire Charles Fabry, 91127 Palaiseau Cedex, France}
\affiliation{Nanomaterials and Nanotechnology Research Center (CINN-CSIC), 
		Universidad de Oviedo (UO), Principado de Asturias, 33940 El Entrego, Spain}	
		
\author{Norman~Y.~Yao}
\affiliation{Department of Physics, Harvard University, Cambridge, Massachusetts 02138 USA}
		
\author{Thierry~Lahaye}
\affiliation{Universit\'e Paris-Saclay, Institut d'Optique Graduate School,\\
		CNRS, Laboratoire Charles Fabry, 91127 Palaiseau Cedex, France}

\author{Fabio~Mezzacapo}
\affiliation{Univ Lyon, Ens de Lyon, CNRS, Laboratoire de Physique, F-69342 Lyon, France}	
		
\author{Tommaso~Roscilde}
\affiliation{Univ Lyon, Ens de Lyon, CNRS, Laboratoire de Physique, F-69342 Lyon, France}	

\author{Antoine~Browaeys}
\affiliation{Universit\'e Paris-Saclay, Institut d'Optique Graduate School,\\
		CNRS, Laboratoire Charles Fabry, 91127 Palaiseau Cedex, France}

\date{\today}

\begin{abstract}
We use a Rydberg quantum simulator to demonstrate a new form of spectroscopy, 
called \emph{quench spectroscopy}, which probes the low-energy excitations of a many-body system.
We illustrate the method on a two-dimensional simulation of the spin-1/2 dipolar XY model. 
Through microscopic measurements of the spatial  spin correlation dynamics following a quench, we extract the 
dispersion relation of the elementary excitations for both ferro- and anti-ferromagnetic couplings. 
We observe qualitatively different behaviors between the two cases that result from the long-range 
nature of the interactions, and the frustration inherent in the antiferromagnet.
In particular, the ferromagnet exhibits elementary excitations behaving as linear spin waves. 
In the anti-ferromagnet, spin waves appear to decay, suggesting the presence of strong nonlinearities.
Our demonstration highlights the importance of 
power-law interactions on the excitation spectrum of a many-body system.
\end{abstract}

\maketitle

The nature and spectrum of elementary excitations are defining features of quantum matter. 
They reflect low-energy physics, and dictate how quantum information 
propagates in the system  \cite{Calabrese2006,Bravyi2006,Cheneau2012}. 
These elementary excitations are often waves
characterized by a dispersion relation $\omega_{\bm k}$ that connects their energy
to their wavevector $\bm k$.
This relation depends on  the dimensionality of the system, its symmetries, 
and the range of the interactions between particles.
Excitations are traditionally probed in condensed matter via linear response, 
which relies on cooling the system  down to low temperatures and measuring its equilibrium 
response to weak perturbations~\cite{Foersterbook,Lovesey1980,Sobota2021,Lovesey1984}.
The ability offered by synthetic quantum systems to monitor (nearly) unitary evolution, 
in both space and time, provides an alternative approach to probe the excitations: one can inject
a finite density of excitations into the system and observe the subsequent dynamics
-- a so-called \emph{quench} experiment~\cite{Mitra2018}. 
Following the quench, the re-organization of spatial correlations is then dictated by the 
propagation of excitations, which is governed by their dispersion relation \cite{Cevolani2018,Despres2019,Schneider2021}. 
This intuition can be put on firm ground when the elementary excitations are free quasi-particles: 
in that case, the Fourier transform  of the equal time correlations in space at wavevector ${\bm k}$
-- hereafter termed the time-dependent structure factor (TSF) -- 
is expected to oscillate in time at frequency $2\omega_{\bm k}$ \cite{Menu2018,Frerot2018,Villa2019,Villa2020,Menu2023}. 
This result forms the basis of \emph{quench spectroscopy}, 
which we implement here.

\begin{figure*}
	\centering
	\includegraphics[width=\linewidth]{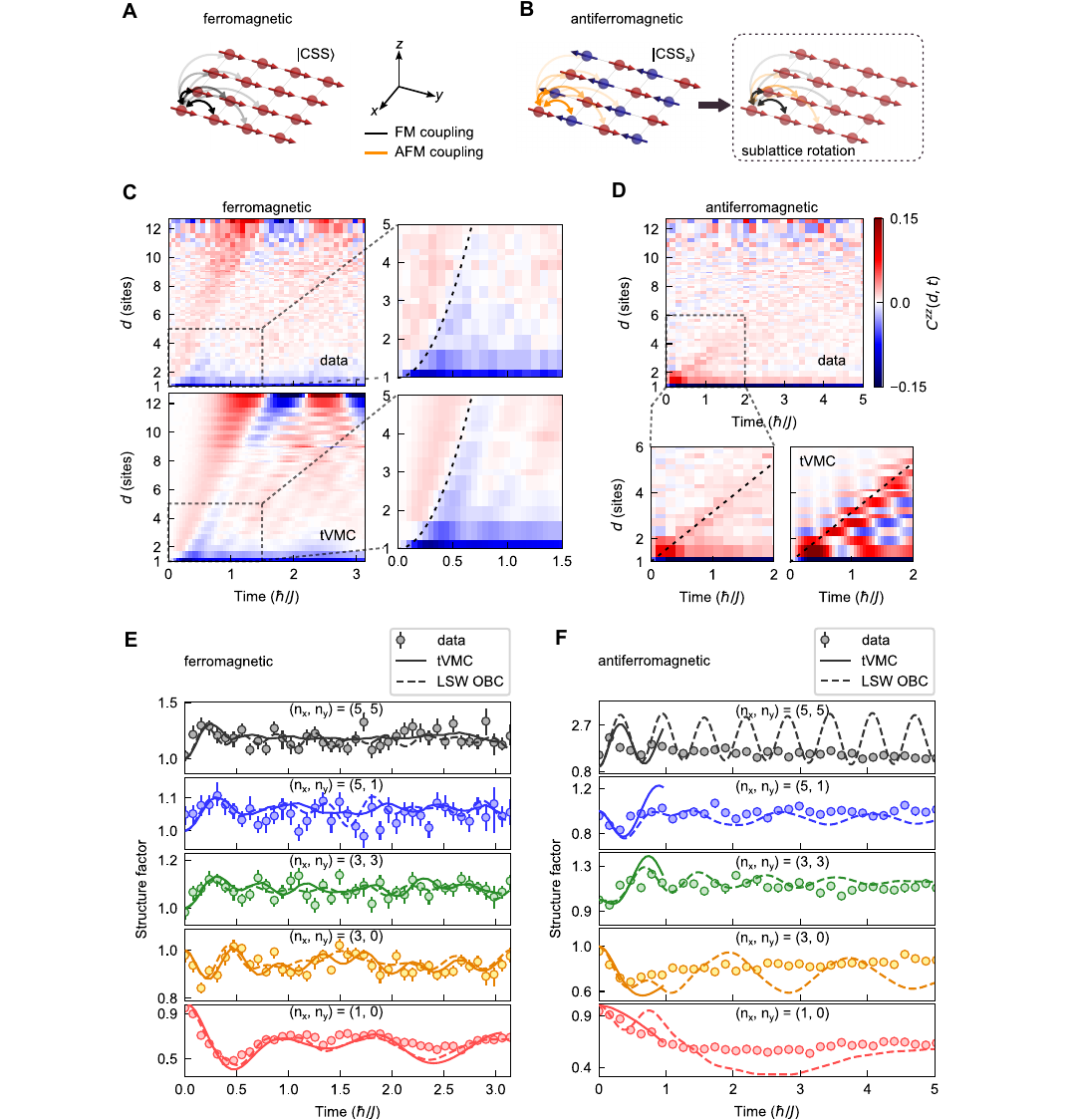}
	\caption{\textbf{Quench spectroscopy of the 2$d$ dipolar XY model.}
		\textbf{(A)} Schematic representation of the classical ferromagnetic state along $y$
		prepared during the quench,
		$|{\rm CSS}\rangle = \left | \rightarrow\cdots \rightarrow \rightarrow\right \rangle_y$,
		featuring the dipolar FM couplings.  
		\textbf{(B)} Schematic representation of the initial classical antiferromagnetic state along $y$,
		$|{\rm CSS}_s\rangle = \left | \leftarrow\rightarrow \cdots \rightarrow \leftarrow\right \rangle_y$,
		featuring the dipolar AFM couplings. Applying a canonical transformation $\sigma_i^{x(y)} \to (-1)^i\sigma_i^{x(y)}$ 
		on the spins of one of the two sublattices of the AFM state transforms it into a 
		FM CSS state with frustrated couplings (see text).
		\textbf{(C)},\textbf{(D)} Space-time dynamics of the spin correlations following the quench into 
		the coherent spin state $|{\rm CSS}\rangle$ (FM case, (C))
		or the staggered coherent spin state state $|{\rm CSS}_s\rangle$ (AFM case, (D)). 
		The dashed line in (C) is a guide to the eye $d\sim t^2$. The one in (D) corresponds
		to $d \approx 2 v_g t/a$, with $v_g$ extracted from the dispersion relation (Fig.~\ref{fig:fig2}B).  
		\textbf{(E)},\textbf{(F)} Time evolution of the FM and AFM structure factors 
		$S(k_x,k_y,t)$ (with $(k_x,k_y) = (2\pi/L)(n_x,n_y)$) extracted from the data in (C) and (D)
		(circles). The error bars represent $68\%$ confidence intervals, which are generated using the bootstrap method.
		Solid lines: results of tVMC simulation. 
		Dashed lines: results of LSW theory including the experimentally-calibrated state detection errors.
		We show in (F) the results of tVMC only for $t \lesssim  \hbar/J$
		as the results are not quantitatively accurate for later times (see \cite{SM}). }
	\label{fig:fig1}
\end{figure*}

Our experiment employs a Rydberg-atom quantum simulator, consisting of a 
two-dimensional square array of $N = L \times L=10\times10$ $^{87}$Rb 
atoms trapped in optical tweezers. 
We encode the pseudo spin-1/2 using a pair of Rydberg states 
$\left|\uparrow\right\rangle = |60S_{1/2}, m_j = 1/2\rangle$ and 
$\left|\downarrow\right\rangle = |60P_{3/2}, m_j = -1/2\rangle$, 
which are coupled by resonant dipole-dipole exchange between the atoms ~\cite{Browaeys2020,SM}.
The system is well-described by the dipolar XY Hamiltonian,
\begin{equation}\label{Eq:HXY}
		H_{\rm XY}= - {\frac{J}{2}}
		\sum_{i < j}  \frac{a^3}{r_{ij}^3}  (\sigma^x_i \sigma^x_j + \sigma^y_i 	\sigma^y_j),
\end{equation}
where $J/h= 0.25~$MHz is the dipolar interaction strength, $\sigma_i^{x,y,z}$ are Pauli matrices, 
$r_{ij}$ is the distance between spins $i$ and $j$, and $a=15~\mu$m is the lattice spacing.
We apply a magnetic field perpendicular to the lattice plane to define the quantization axis 
and ensure isotropic dipolar interactions.
After initializing the system in an uncorrelated product state, corresponding to 
the mean-field (MF) ground-state for ferromagnetic (FM) or antiferromagnetic (AFM) interactions,  we
then monitor the evolution of equal-time spin correlations in real space,
as well as in momentum space by evaluating the TSF. 
Our main results are twofold.
In the case of FM dipolar interactions, the excitations are found to behave as long-lived
spin waves (also called magnons), whose frequency can be extracted from a fit of the dynamics of the TSF.
We observe a non-linear dispersion relation, in agreement with  the prediction for 
dipolar spin waves in two spatial dimensions where $\omega_{\bm k}\sim \sqrt{k}$ for small $k$ \cite{Peter2012}.
For AFM interactions, a different picture emerges. 
There, the frustrated nature of the antiferromagnetic dipolar interactions 
leads to a linear dispersion relation. 
Moreover, the oscillations of the AFM TSF are strongly damped, suggesting
the existence of decay processes which render spin waves unstable.

We begin the quench spectroscopy protocol by preparing either a uniform 
coherent spin state (CSS) $|{\rm CSS}\rangle = \left | \rightarrow\cdots \rightarrow \rightarrow\right \rangle_y$  
aligned along the $y$ axis, or a staggered CSS  
$|{\rm CSS}_s\rangle = \left | \leftarrow\rightarrow \cdots \rightarrow \leftarrow\right \rangle_y$  (Fig.~\ref{fig:fig1}A,B). 
The uniform CSS is the MF ground state of the FM dipolar XY Hamiltonian 
$H_{\rm XY}$, and its dynamical evolution is thus governed by 
the low-energy excitations above the true ground state.
The staggered CSS, on the other hand, is the MF ground state for AFM interactions, 
namely for $-H_{\rm XY}$. 
Due to time-reversal symmetry, the ensuing evolution can be viewed as low-energy dynamics 
governed by the AFM dipolar XY Hamiltonian $-H_{\rm XY}$ \cite{Frerot2018}.  
To clarify the contrast between the two cases, we apply a canonical transformation in the AFM case, 
rotating spins on one sublattice by $\pi$ around the $z$-axis, 
so that  $\sigma_i^{x(y)} \to (-1)^i \sigma_i^{x(y)}$. 
In this relabeled coordinate system, the initial state 
becomes the same for the FM and the AFM, but the Hamiltonian of the AFM transforms 
into that of a frustrated ferromagnet with staggered couplings, $(J a^3/r_{ij}^3) (-1)^{i+j+1}$ 
(Fig.~\ref{fig:fig1}B).

After a given evolution time $t$, we read out the state of each atom \cite{Chen2023,Bornet2023} 
in the $z$-basis, and reconstruct the equal time correlation function for the $z$ spin components 
of the atom pair $(i,j)$, 
$C^{zz}_{ij}(t) = \langle \sigma^{z}_i \sigma^{z}_j \rangle(t) - \langle \sigma^{z}_i \rangle(t)  \langle \sigma^{z}_j \rangle(t)$.
The choice of the $z$ basis is dictated by the fact that $\sigma^z$-operators correspond to spin flips in the $xy$ plane, 
thus revealing excitations above low-energy states which exhibit long-range order for the 
$x$ and $y$ spin components \cite{SM}.
The space-time evolution of the correlations is presented in Fig.~\ref{fig:fig1}C, D, 
where we show the correlation function averaged over all $N_d$ pairs of spins separated by the same distance 
$d$, $C^{zz}(d,t) =  N_d^{-1} \sum_{ij | r_{ij}=d}  C^{zz}_{ij}(t)$.
Strikingly, the correlations differ between the FM and the AFM at both short and long distances.
At short distances, the correlations are negative in the FM, while they alternate in sign 
(and are stronger) in the AFM -- a subtle effect that we discuss at the end of this paper. 
Moreover, the correlations of the FM  propagate rapidly in space and reach 
the longest-distances in less than an interaction time $\hbar/J$.
The data are also consistent with the super-ballistic propagation of  correlation  fringes
predicted to scale as $d \sim t^2$  for the dipolar XY model~\cite{Peter2012,Frerot2018}.
This is highlighted by the dashed line in Fig.~\ref{fig:fig1}C.  
In the AFM, we find that the correlations propagate slower, and that the correlation 
front moves linearly with time (dashed line in Fig.~\ref{fig:fig1}D). 
This difference is due to frustration, which effectively shortens 
the range of the interaction and leads to a finite maximal group velocity 
$v_g = \max_{\bm k} |\nabla_{\bm k}\omega_{\bm k}| \approx J a/\hbar$ 
of the elementary excitations, such that $d \approx 2 v_g t/a$ \cite{Cheneau2012}. 
Moreover, the correlations appear to vanish at a finite distance of $\sim$ 6 sites, 
a consequence of thermalization towards a state with  short-range correlations (see below).

\begin{figure*}
	\centering
	\includegraphics[width=\linewidth]{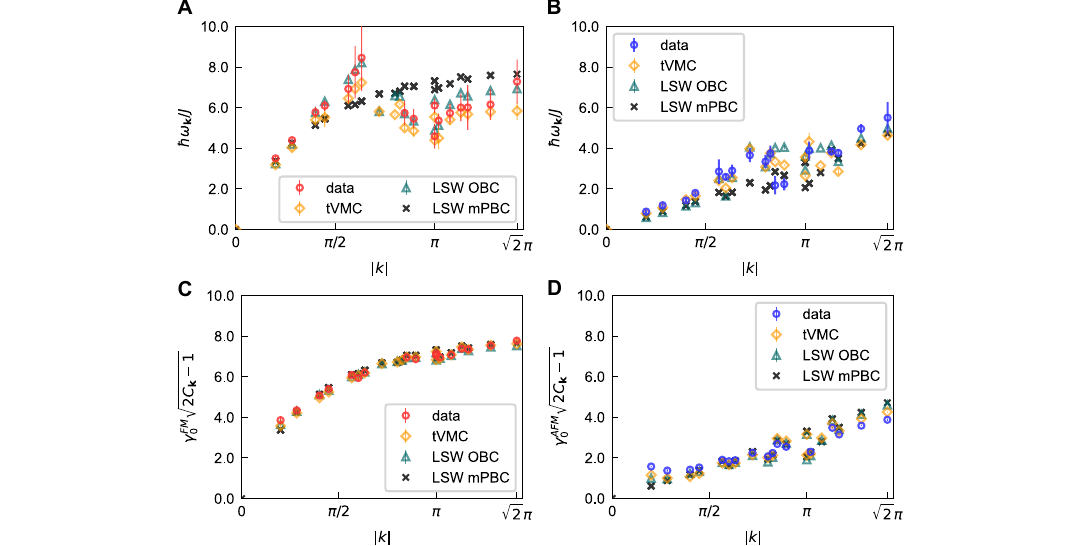}
	\caption{\textbf{Measurement of the dispersion relation of the dipolar XY model.} 
	\textbf{(A)},\textbf{(B)} Frequency $\omega_{\bm k}$ as a function of $|{\bm k}| = \sqrt{k_x^2+k_y^2}$, 
	fitted from the time-dependent structure factors shown in Fig. \ref{fig:fig1}E,F.
	The ferromagnetic case (A) exhibits a non-linear dispersion relation.
	The antiferromagnetic case  (B) shows a linear dispersion relation.
	\textbf{(C)},\textbf{(D)} Frequencies $\omega_{\bm k}$ obtained by using Eq.~(\ref{Eq:disp_rel_Cfit})
	and the fitted offset of $S(\bm k,t)$. 
	The data (circles) are compared to the results of LSW  with mPBC (crosses) and tVMC (diamonds), see text. 
	On all figures, the error bar represents the standard error from the fits. 
	}\label{fig:fig2}
\end{figure*}

The TSF is then obtained by a spatial Fourier transform of the equal-time correlation function: 
\begin{equation}
S(\bm k,t)= (1/N) \sum_{i,j} e^{i{\bm k}\cdot {\bm r}_{ij}} C^{zz}_{ij}(t),
\end{equation}
with wave-vectors $\bm k = (2\pi/L)(n_x,n_y)$ where $n_{x(y)} = -L/2+1,..., L/2$. 
As observed in Fig.~\ref{fig:fig1}E,F, the TSF exhibits oscillations at short times for all wavevectors
(full data set in \cite{SM}). 
In the FM case, the oscillations persist at longer times for small wave-vectors. 
The dynamics of the AFM are slower than its FM counterpart at the same wavevector,
and appear damped at long times. 
In both cases, we compare the experimental data with the predictions of 
time-dependent linear spin-wave (LSW) theory 
including a constraint due to finite-size (see below)
and with time-dependent variational Monte Carlo (tVMC) (see details in \cite{SM}). 
In the FM case, we observe  good agreement between the experimental results and both 
theoretical predictions (which also agree with one another), suggesting that the elementary excitations 
of the FM  behave as undamped spin waves. 
Contrarily, the AFM data are in rather poor agreement with LSW theory, which does not 
predict any damping -- thus suggesting that the AFM dynamics is strongly non-linear.

We then extract the dispersion relation $\omega_{\bm k}$ of elementary excitations from a fit of $S(\bm k, t)$ 
to the form $ A_{\bm k} \cos(2\omega_{\bm k} t +\phi_{\bm k}) + C_{\bm k}$, predicted by LSW theory.
Although our system lacks well-defined wavevectors due to its open boundary conditions (OBC), 
this fit is still applicable because its early-time dynamics is approximately equivalent to that of a 
system with periodic boundary conditions and modified couplings (mPBC), as we detail in \cite{SM}. 
We thus restrict the fits to early times, \emph{i.e.} to the data points corresponding typically to the first oscillation. 
In Fig.~\ref{fig:fig2}, we show the dispersion relation $\omega_{\bm k}$ extracted from fits to the 
experimental data, to the tVMC simulation, and to the LSW calculations. We 
also compare to the predictions of LSW with mPBC~\cite{Peter2012,Frerot2018}:
for dipolar XY systems with mPBC, one expects 
$\hbar \omega_{\bm k}/J = \gamma_0 \sqrt{1- \gamma_{\bm k}/ \gamma_0}$, 
where $\gamma_{\bm k} = (1/N)\sum_{i\neq j} e^{i {\bm k} \cdot {\bm r}_{ij}} \eta^{i+j+1} a^3/r_{ij}^3$ 
with $\eta = 1 (-1)$ for the FM (AFM).
In the FM case, we observe a non-linear dispersion down to the smallest wavevector.  
The extracted dispersion agrees well with tVMC, and closely follows the LSW prediction 
that $\omega_{\bm k} \sim \sqrt{k}$ as $k\to 0$  due to the long-range dipolar tail.
For the AFM, we instead  obtain  a linear dispersion relation at small $k$ (and even beyond), 
characteristic of the effective short-range interactions induced by frustration~\cite{Frerot2017}.
This is also in agreement with the light-cone picture of correlation spreading in 
real space seen in Fig.~\ref{fig:fig1}, with a spin-wave velocity $v_g \approx J a/\hbar$.

However, while LSW theory for FM matches the experiment rather accurately at small $k$, 
systematic deviations appear at larger wavevectors, which we also observe on the tVMC data and LSW with OBC. 
This is a finite-size effect, whose origin can be traced to the local constraint $\langle(\sigma_i^z)^2\rangle = 1$ 
for spin-1/2 systems, which imposes a TSF sum rule beyond standard LSW theory \cite{SM}.
Enforcing this sum rule alters the frequency 
of the short-time behavior with respect to the LSW prediction, for any system size. 
As we show in \cite{SM}, LSW theory predictions would be recovered at long times for systems larger than ours.
Nonetheless, there exists an alternate prediction from LSW theory which is minimally 
affected by this nonlinear constraint, even in our finite-size system \cite{SM}.  
Namely,  the offset $C_{\bm k}$ is related to the frequency $\omega_{\bm k}$ as \cite{Frerot2018}:
\begin{equation}\label{Eq:disp_rel_Cfit}
		\hbar \omega_{\bm k} = J \gamma_0 \sqrt{2C_{\bm k}-1}\ ,
\end{equation}
where $\gamma_{0}^{\textrm{FM}} \approx 6.5$ and $\gamma_{0}^{\textrm{AFM}} \approx 2.5$ 
for our $10\times10$ array.
We extract $C_{\bm k}$ from a fit to the data shown in Fig.\,\ref{fig:fig1}E, F, 
and infer $\omega_{\bm k}$ using Eq.\,(\ref{Eq:disp_rel_Cfit}). 
The thus-inferred dispersion relation is shown in Fig.~\ref{fig:fig2}C,D, for both the experimental and tVMC data. 
We also compare to the predictions of LSW theory for a system with mPBC. 
In the FM, we observe excellent agreement, which confirms the  validity of LSW theory for this system. 
For the AFM the overall agreement is also good, but systematic deviations are observed, 
related to the fact that spin waves are not stable (see below), and possibly to the imperfections
in the initial state preparation, larger than for the FM \cite{SM}.
	
\begin{figure*}
\centering
	\includegraphics[width=\linewidth]{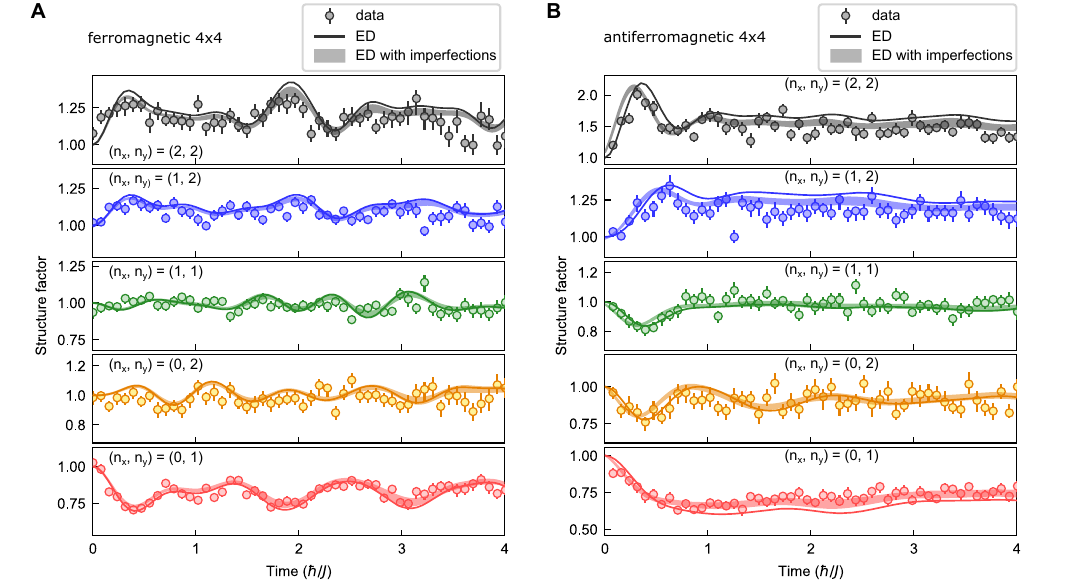}
	\caption{\textbf{Time evolution of the structure factors $S(k_x,k_y,t)$ for a 4$\times$4 array.} 
	The solid lines are the results of exact diagonalization simulations, without or with experimental imperfections. 
	Comparing the two, we conclude that the motion of the atoms and finite Rydberg lifetime contribute to the increase 
	of the error bars at late times (see Supplementary Materials~\ref{SM:exp_imperfections}). 
	The width of the shaded region corresponds to a 1-$\sigma$ standard 
	deviation obtained from 100 realizations including imperfections. 
	\textbf{(A)},\textbf{(B)} ferromagnetic and antiferromagnetic case, respectively. 
	The experimental error bars represent $68\%$ confidence intervals, which are generated using the bootstrap method.}\label{fig:fig3}
\end{figure*}

We now address the long-time dynamics, which, as previously noted, differs for the FM and AFM.
First, to assess whether the decay observed in Fig.~\ref{fig:fig1}F for the AFM dynamics is 
due to experimental imperfections, or if it is intrinsic to the unitary dynamics of the system, 
we scale our experiment down to $N=4\times 4$ atoms, which can be exactly simulated, 
including all the known experimental imperfections (details of imperfections in SM \cite{SM}). 
The results are presented in Fig.~\ref{fig:fig3}A,B. 
In both the FM and the AFM, we observe that the experimental imperfections do not 
significantly alter the dynamics at long times.
This suggests that the decay of oscillations in the AFM is indeed intrinsic, 
which may be attributed to an instability of the spin-wave excitations. 
As we further elaborate in the SM \cite{SM},  the origin of this instability can be  attributed to several effects.
The first non-linear correction to the AFM spin-wave theory predicts a decay of
single magnons into three magnons of lower energy and momentum. 
This decay is instead negligible for the FM, due to kinematical constraints on magnon 
decay that lead to a lack of phase space for the decay process.  
Moreover, as revealed by a systematic study of different spectral functions (see \cite{SM}), 
the decay must also come from multi-magnon processes resulting from 
the finite density of magnons injected into the system during the initial state preparation.

Second, we discuss the issue of thermalization at long times. 
For both the FM and the AFM, one could expect equilibration of local 
observables to thermal equilibrium at a temperature  $T_{\rm CSS}$ 
corresponding to the energy of the initial state, namely such that 
$\langle H_{\rm XY} \rangle_{T_{\rm CSS}} = \langle {\rm CSS} |  H_{\rm XY} | {\rm CSS}\rangle$. 
Experimentally, after the short initial transient from which we extract the dispersion relation, 
the TSF oscillates around a  well-defined value, which can be interpreted as  
the onset of equilibration.
Assuming the eigenstate-thermalization hypothesis~\cite{DAlessio2016}, these 
oscillations should take place around the thermal-equilibrium value of the TSF 
at $T_{\rm CSS}$, with an amplitude decreasing exponentially with system size. 
In the FM, however, this prediction should be taken with some caution, 
as the dynamics of linear spin waves has an effectively integrable nature. 
As such, it cannot lead to proper thermalization, but only to dephasing 
(or pre-thermalization \cite{Mori2018}) when looking at quantities which probe several spin-wave frequencies. 
For the AFM, instead, the damping of oscillations suggests a fully chaotic dynamics and proper thermalization. 

\begin{figure*}
	\centering
	\includegraphics[width=\linewidth]{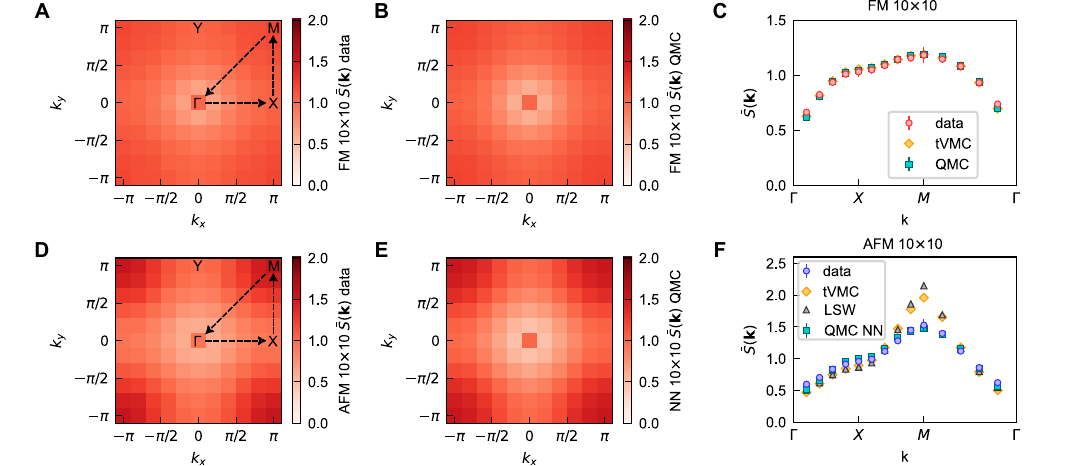}
	\caption{\textbf{Time-averaged structure factor $\bar{S}(\bm k)$.}
	\textbf{(A)},\textbf{(B)} (\textbf{(D)},\textbf{(E)}) Comparison of experimental data and Quantum Monte Carlo 
	simulations for the ferromagnetic (antiferromagnetic) case, plotted in the first Brillouin zone. 
	\textbf{(C)},\textbf{(F)} Experimental time-average structure factor $\bar{S}(\bm k)$ plotted along the path connecting 
	the high-symmetry points of the Brillouin zone (dotted lines in (A),(D)). We also plot the predictions
	from tVMC, Quantum Monte Carlo (FM case), LSW (AFM) and QMC with nearest-neighbor (NN) interactions (AFM).}\label{fig:fig4}
\end{figure*}

For both the FM and  AFM we extract the equilibrium value of the TSF by averaging 
the data after the initial transient of duration $\hbar/J$. 
The resulting time-averaged structure factors $\bar{S}(\bm k)$ are shown in Fig.~\ref{fig:fig4}. 
For the FM, we calculate the thermalized structure factor via equilibrium quantum 
Monte Carlo (QMC) \cite{SM} (accounting for the conservation of structure factor at $k=0$). 
We find that this value is in excellent agreement with the one averaged over the 
oscillations, indicating that the pre-thermalized structure factor coincides with the thermal one.
The corresponding temperature is $T_{\rm CSS}^{\rm FM}\approx 1.2 J/k_{\rm B}$, lying deep inside 
the long-range ordered regime~\cite{sbierski2024,SM} where spin waves should be the 
dominant thermally populated excitations. 
We also find that their density remains small (see SM \cite{SM}), justifying the picture of linear dynamics. 

In the AFM case, predicting the thermal value for the TSF is challenging due to the sign problem in QMC. 
Using tensor network methods, we estimate that $T_{\rm CSS}^{\rm AFM} \approx 0.6 J/k_{\rm B}$, 
falling in a paramagnetic phase above a Berezinskii-Kosterlitz-Thouless transition \cite{Chen2023,SM}. 
In this regime, non-linear excitations  (including unbound vortex-antivortex pairs) should be thermally populated. 
This is reflected by a significantly higher density of spin waves 
developed during the dynamics~\cite{SM}, which implies non-linearities play a prominent role.
Nonetheless, we again observe a relatively good agreement between 
the experimental data and the  predictions of tVMC and LSW theory. 
For comparison, we also plot the results of an equilibrium QMC calculation for the XY model with 
nearest-neighbor interactions, at the corresponding $T_{\rm CSS}$ for that model.
The agreement with the experimental data is unexpectedly good, 
suggesting again the effective short-range nature of the frustrated dipolar interactions.

Both the FM and the AFM show a peak of $\bar{S}(\bm k)$ at ${\bm k} a= (\pi,\pi)$, 
reflecting the short-range anti-correlations that we observed  in Fig.~\ref{fig:fig1}C, D. 
To understand the structure of these correlations 
emerging at long times, we map the dipolar spin-1/2 XY 
model onto hardcore bosons \cite{Matsubara1956}: 
the anti-correlations in the $\sigma^z$ spin components
reflects the tendency of hardcore bosons to form  staggered density patterns 
for which the dipolar hoppings allow them to delocalize  and reduce their kinetic energy. 
This has been observed in one-dimensional repulsive bosonic gases, 
leading to short-range crystallization \cite{Meinert2017}. 
Surprisingly the frustrated hoppings of the AFM case appear to lead to 
significantly stronger correlations (namely, a higher peak) than the unfrustrated, FM ones. 
This behavior is opposite to what happens to phase correlations -- 
namely the correlations among the $\sigma^{x(y)}$ spin components -- 
which thermalize to long-range order in the FM case, 
while they are expected to thermalize to short-range order in the AFM case due to frustration \cite{Chen2023}. 
This enhancement of $\sigma^z$ correlations by frustrating the ordering of the 
$\sigma^{x(y)}$ components can be understood as a result of the effective hopping 
range in the corresponding bosonic model: In the FM case, bosons delocalize and 
correlate in phase over all distances thanks to the dipolar hopping, 
so that short-range density correlations are weakened; 
contrarily, in the AFM case, bosons delocalize only at short range 
because of the frustrated hopping, and they do so by enhancing (short-range) density correlations. 

In conclusion, we have experimentally reconstructed  the dispersion relation of 
a strongly interacting quantum system using quench spectroscopy. 
Our results highlight the special role of ferromagnetic dipolar interactions in two dimensions, 
whose long-range tail leads to a non-linear dispersion relation of excitations. 
The frustration resulting from antiferromagnetic dipolar interactions instead leads  to 
an effective cancellation of the long-range tail, recovering a linear dispersion relation 
as in the case of finite-range interactions.  
We observe that the elementary excitations of the dipolar ferromagnet have 
the nature of linear spin waves, leading to effectively integrable dynamics at low energy. 
On the contrary, the dipolar antiferromagnet shows a decay of spin-wave oscillations,
highlighting the importance of non-linear quantum fluctuations in the system.  
Our work  moreover demonstrates the need to account for local constraints when analyzing quench dynamics.
Their effects, which would vanish at long time for larger system sizes, should  
be visible in all quantum simulation platforms as they operate at about the same size as ours.

The quench spectroscopy demonstrated here can be applied more broadly.  
A particular advantage of this method compared  to other approaches 
\cite{Hild2014,Jepsen2020,GuardadoSanchez2020,Morvan2022,Jurcevic2014,Kranzl2023} 
is a reduced need for local control: 
it relies only on initialization into the mean-field ground state of the Hamiltonian, 
in order to target the low-energy spectrum \cite{Villa2019,Villa2020}.  
Similar global quenches have been adopted in previous experiments on 
dilute Bose gases~\cite{Hung2013,Schemmer2018} and trapped ion chains~\cite{tan2021domain}. 
Looking forward, our procedure can be extended by using different initial states in order to 
address higher-energy regions of the excitation spectrum. 
It could also be applied to more exotic phases of matter,  such as magnetism on 
frustrated triangular or kagom\'e lattices,  supporting strongly fluctuating ordered 
phases or perhaps even fractionalized spin liquids \cite{Yao2018}. 

\begin{acknowledgments}
We acknowledge insightful discussions with H.P. B\"uchler, L. Sanchez-Palencia, 
M. Zaletel, C. Laumann and M. Schuler. 
This work is supported by
the Agence Nationale de la Recherche (ANR, project RYBOTIN and ANR-22-PETQ-0004 France 2030, project QuBitAF),
 Horizon Europe programme HORIZON-CL4-2022-QUANTUM-02-SGA via the project 101113690 (PASQuanS2.1),
and the European Research Council (Advanced grant No. 101018511-ATARAXIA). 
All numerical simulations were performed on the PSMN cluster at the ENS Lyon.
N.Y.Y. acknowledges support from the Army Research Office (ARO) 
(grant no. W911NF-21-1-0262) and a Simon's Investigator award. 
M.B. acknowledges support from the NSF through the QLCI programme 
(grant no. OMA-2016245) and the Center for Ultracold Atoms. 
V.L. acknowledges support from the Wellcome Leap as part of the Quantum for Bio Program.
S.C. acknowledges support from the U.S. Department of Energy, Office of Science, 
through the Quantum Systems Accelerator (QSA), 
a National Quantum Information Science Research Center.
D.B. acknowledges support from MCIN/AEI/10.13039/501100011033 
(RYC2018-025348-I, PID2020-119667GA-I00 and European Union NextGenerationEU PRTR-C17.I1)
\end{acknowledgments}

\bibliography{reference_dispersion_relation}

\clearpage

\setcounter{figure}{0}
\renewcommand\thefigure{S\arabic{figure}} 



%

\begin{center}
{\bf Supplemental Material}
\end{center}

\section{Experimental methods} 

\subsection{Experimental procedures}\label{SM:Exp_details}

\begin{figure*}
\includegraphics[width=\linewidth]{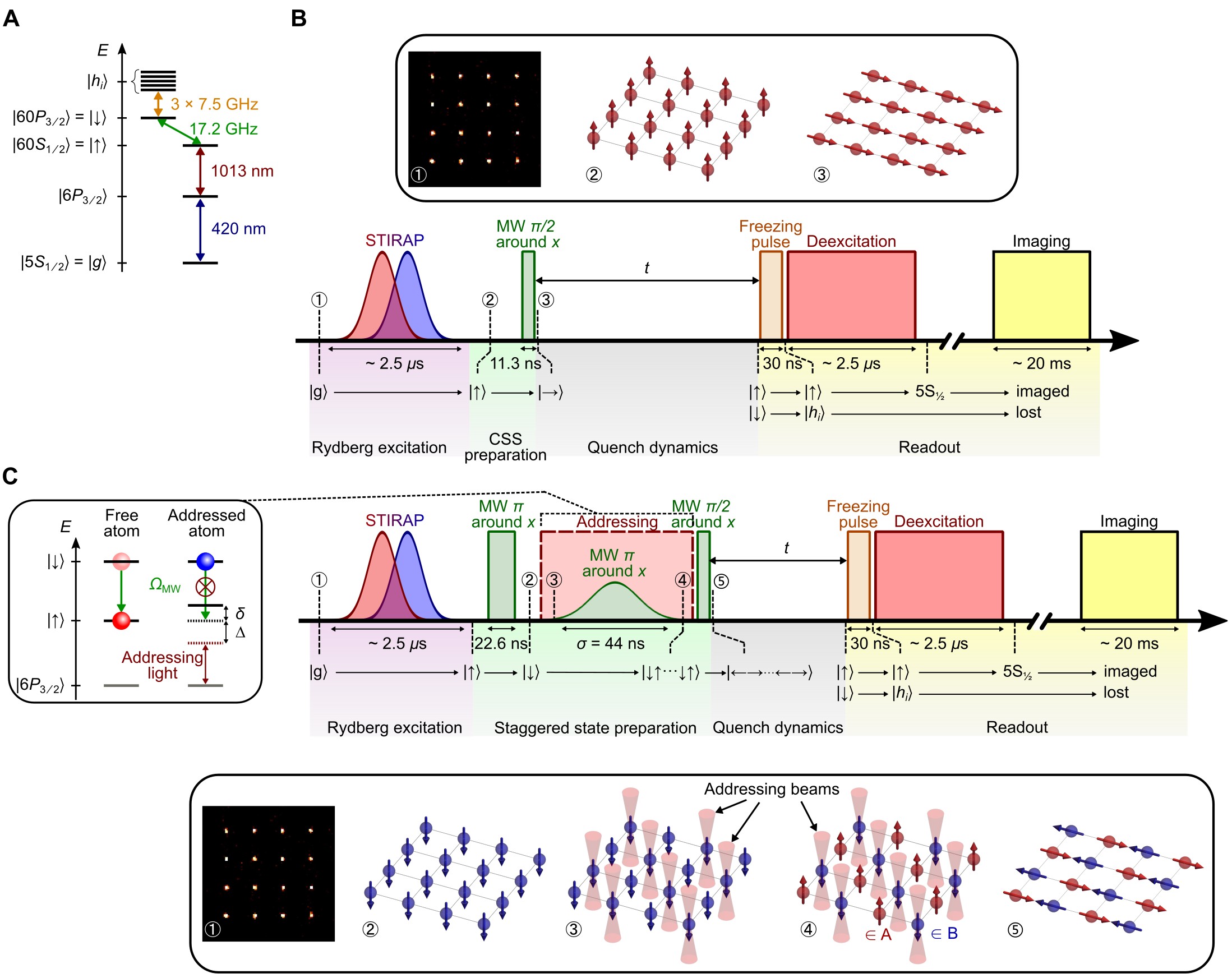}
\caption{\textbf{Experimental sequence.}
\textbf{(A)}~Schematics of the atomic levels relevant for the experiment.
\textbf{(B)}~Sequence of optical and microwave pulses (not to scale) used for 
the experiments in the ferromagnetic case reported in the main text.
\textbf{(C)}~Sequence of optical and microwave pulses (not to scale) used for 
the experiments in the antiferromagnetic case reported in the main text.}
\label{fig:ExpSequ_SM}
\end{figure*}

The implementation of the dipolar XY Hamiltonian is based on the $^{87}\text{Rb}$ 
Rydberg-atom tweezer array platform described in previous works~\cite{Chen2023, Bornet2023}. 
The pseudo spin-states $\ket{\uparrow} = \ket{60S_{1/2}, m_J = 1/2}$ 
and $\ket{\downarrow} = \ket{60P_{3/2}, m_J = -1/2}$ can be coupled 
by microwave at 17.2~GHz (see Fig.~\ref{fig:ExpSequ_SM}A). 
To isolate the $\ket{\uparrow}-\ket{\downarrow}$ 
transition from irrelevant Zeeman sublevels we apply a $\sim45$-G 
quantization magnetic field perpendicular to the array.

\emph{Experimental sequence} (Fig.~\ref{fig:ExpSequ_SM}) -- 
We assemble arrays of atoms 
trapped in 1-mK deep optical tweezers~\cite{Barredo2016}, 
Raman sideband cool them down to $10\,\mu$K and optically 
pump them into $\ket{g}=\ket{5S_{1/2}, F = 2, m_F = 2}$. 
We then adiabatically reduce the trap depth by a factor $\sim 50$ to further reduce the temperature. 
The tweezers are finally switched off and the atoms 
are excited to the Rydberg state $\ket{\uparrow}$
by a stimulated Raman adiabatic passage (STIRAP) 
with 421-nm and 1013-nm lasers.

To initialize the atoms in $|{\rm CSS}\rangle= \ket{\rightarrow \cdots \rightarrow}$ (FM case), 
we apply a global resonant microwave $\pi/2$ pulse around $x$, 
with a Rabi frequency $\Omega = 2\pi \times 22.2$~MHz (Fig.~\ref{fig:ExpSequ_SM}B). 
For the AFM case, we need to initialize the system in 
$|{\rm CSS}_s\rangle = \left | \leftarrow\rightarrow \cdots \rightarrow \leftarrow\right \rangle_y$. 
As the microwave field does not allow for local manipulations, we combine them with local 
addressing laser beams and use the following procedure. 
We  first apply a global resonant microwave $\pi$ pulse around $x$ 
transferring all the spins from $\ket{\uparrow}$ to $\ket{\downarrow}$, 
(Fig.~\ref{fig:ExpSequ_SM}C).
We then apply a staggered addressing light-shift ($\sim 11$~MHz) on half of the 
atoms and, simultaneously, a weaker microwave $\pi$-pulse ($\Omega = 2\pi \times 7.7$~MHz) 
to transfer only the non-addressed atoms back to $\ket{\uparrow}$ while keeping 
the addressed atoms in $\ket{\downarrow}$. This leads to the AFM state along $z$
$\left | \downarrow\uparrow \cdots \downarrow \uparrow\right \rangle$. 
Finally, we apply a global resonant microwave $\pi/2$ pulse around $x$ 
($\Omega = 2\pi \times 22.2$~MHz) to get to the state $|{\rm CSS}_s\rangle$.

\emph{State-detection procedure} -- The detection protocol includes three parts.
First, we apply a $7.5~$GHz microwave pulse (``freezing pulse'' in 
Fig.~\ref{fig:ExpSequ_SM}) to transfer the population from $\ket{\downarrow}$ 
to the $n=58$ hydrogenic manifold (states $\ket{h_i}$ in Fig.\,\ref{fig:ExpSequ_SM}A). Atoms in the 
hydrogenic states are decoupled from those in $\ket{\uparrow}$, thus avoiding 
detrimental effects of interactions during the read-out sequence.
Second, a $2.5~\mu$s-deexcitation pulse resonant with the transition between 
$\ket{\uparrow}$ and the short-lived intermediate state $6P_{3/2}$ leads to
decay of the atoms back to $5S_{1/2}$.
Third, we switch the tweezers back on to recapture and image (via fluorescence) 
\emph{only} the atoms in $5S_{1/2}$ (the others being lost). 
This protocol maps the $\ket{\uparrow}$ (resp. $\ket{\downarrow}$) state to the presence 
(resp. absence) of the corresponding atom.
The experimental sequence is typically repeated $400$ times with defect-free 
$4\times4$ assembled  arrays, and  $\sim400$ shots with at most 2 defects allowed 
in the $10\times10$ assembled arrays. This allows us to reconstruct the spin correlations 
and the structure factors by averaging over these repeated measurements.

\subsection{Experimental imperfections}\label{SM:exp_imperfections}

Several sources of state preparation and measurement (SPAM) error 
contribute to affecting the observed structure factors. 

\emph{State preparation errors} -- We estimate that the Rydberg excitation process 
is $98\%$ efficient: a fraction $\eta = 2\%$ of the atoms remains in the state 
$|g\rangle$ after Rydberg excitation and hence do not participate in the dynamics. 
These uninitialized atoms are read as a spin $\left\vert \uparrow \right\rangle$ at the end of the sequence.
For the FM case, the following $\pi/2$-microwave pulse used to prepare CSS is also imperfect 
due to the unavoidable effect of 
the dipolar interactions between the atoms during its application: it reduces the 
initial polarization by $\sim 1\%$.
For the AFM case, the $\pi$-microwave pulse applied simultaneously to the addressing 
laser has an efficiency $ \eta_{\text{non-add}}$ to transfer 
the non-addressed atoms back to $\ket{\uparrow}$, and $\eta_{\text{add}}$ 
to transfer the addressed atoms to $\ket{\uparrow}$. 
For the $4\times4$ array with $\sim 68~$MHz addressing light-shift we find
$ \eta_{\text{non-add}} = 98\%$ and $\eta_{\text{add}} = 2\%$.
For the $10\times10$ arrays with $\sim 11~$MHz addressing light-shift, we get
$\eta_{\text{non-add}} = 95\%$ and $\eta_{\text{add}} = 2\%$. The fact that $\eta_{\text{add}}$
has the same value in the two cases is fortuitous and indicates that processes other than off-resonant 
microwave transitions play a role.

\emph{Measurement errors} -- 
Due to the finite efficiency of each step in the readout sequence (see Fig.~\ref{SM:Exp_details}B), 
an atom in $\ket{\uparrow}$ (resp. $\ket{\downarrow}$) 
has a non-zero probability $\epsilon_{\uparrow}$ (resp. $\epsilon_{\downarrow}$) 
to be detected in the wrong state~\cite{Chen2023}. 
The main contributions to $\epsilon_{\uparrow}$ are the finite efficiency $1-\eta_{\text{dx}}$ 
of the deexcitation pulse and the probability of loss $\epsilon$ due to collisions with the background gas. 
As for $\epsilon_{\downarrow}$, the main contribution is the $\ket{\downarrow}$ Rydberg state radiative lifetime.
A set of calibrations leads, to first order, to $\epsilon_{\uparrow} \simeq \eta_{\text{dx}} + \epsilon = 1.5\% + 1.0\% = 2.5\%$ 
and $\epsilon_{\downarrow} = 1.0\%$.

The experimental structure factors $S(k_x,k_y;t)$ are related to the same quantities 
$\tilde{S}(k_x,k_y;t)$ {\it without} detection errors by the following equations 
(valid to first order in $\epsilon_{\uparrow,\downarrow}$):
\begin{equation}\label{Eq:correction}
\begin{split}
S(k_x,k_y;t) =~& (1-2\epsilon_{\downarrow}-2\epsilon_{\uparrow})\tilde{S}(k_x,k_y;t) \\ & + 
2\epsilon_{\downarrow}+ 2\epsilon_{\uparrow}.
\end{split}
\end{equation}
The results of numerical simulations and spin-wave theory shown in the 
various figures include the detection errors. 

\emph{Decoherence} -- It comes from two effects.
First, the atomic spacings exhibit shot-to-shot fluctuations due to the positions and 
velocities distributions of the atoms in the trap. We consider that they 
follow Gaussian distributions with standard deviations $\sigma_{x/y} \sim 100~$nm 
in the array plane, $\sigma_{z} \sim 800~$nm and $\sigma_{v_x/v_y}\sim 0.025~\mu\text{m}/\mu$s 
in radial directions, $\sigma_{v_z}\sim 0.04~\mu\text{m}/\mu$s in axial directions. 
The resulting fluctuations of the atomic interaction leads to 
decoherence when averaging over many realizations.
Second, a Rydberg atom can decay spontaneously to the ground state
or be transferred by black-body radiation to other Rydberg states. 
When this occurs, we consider the atom loses the coupling with the other atoms, 
which affects the ensuing dynamics.
The lifetimes at 300~K of the 60S and 60P state are 98~$\mu$s and 120~$\mu$s, respectively. 
These two time-dependent factors contribute to the increase in the error bars at long time in the simulations
shown in Fig.~\ref{fig:fig3}.

\section{Representative fits of the experimental data for the time-dependent structure factor}

\begin{center}
\begin{figure*}
\includegraphics[width=\textwidth]{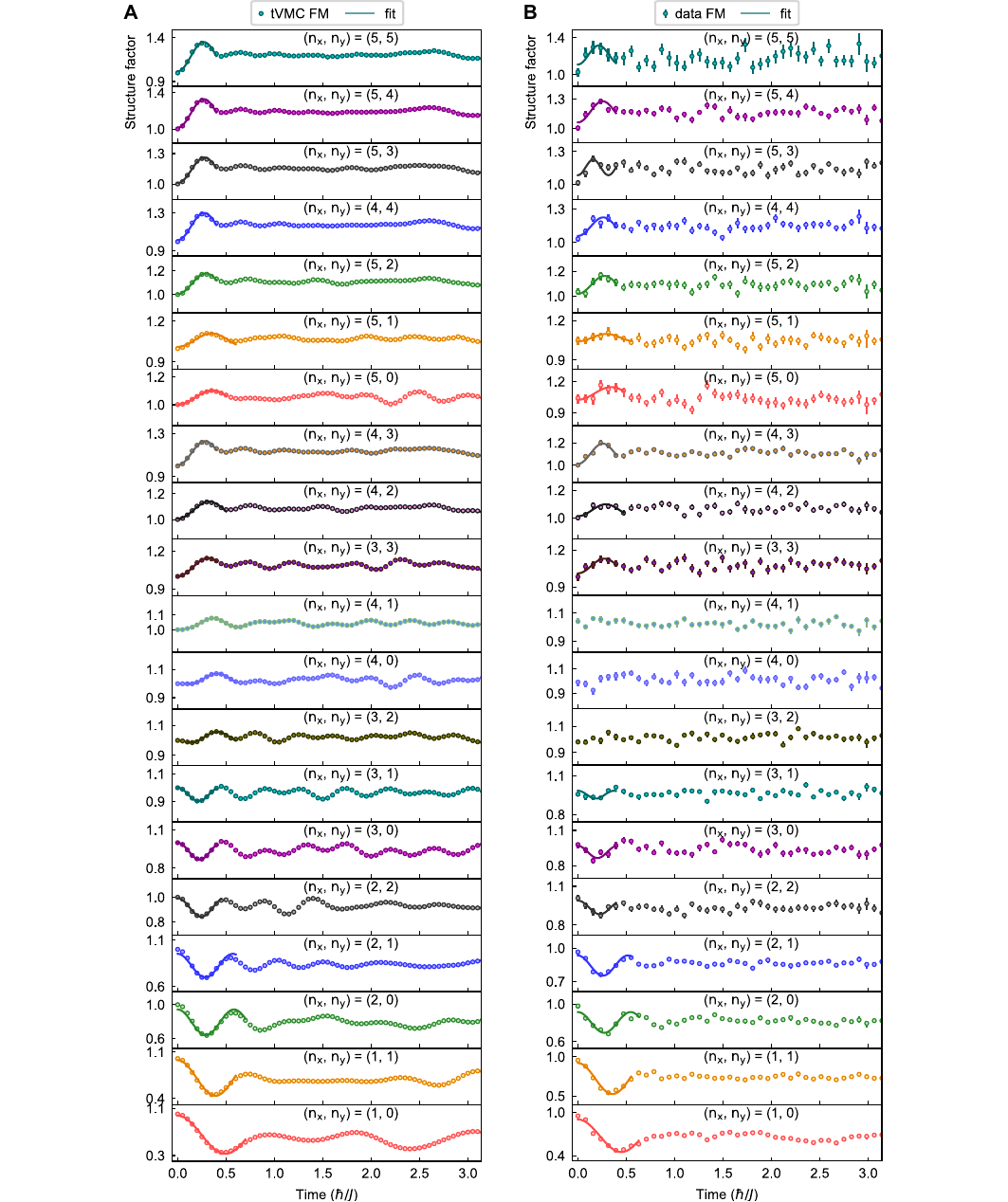}
\caption{\textbf{Full data sets for the time-dependent ferromagnetic structure factor $S(k_x,k_y;t)$}. 
(Left): tVMC calculations. (Right): experimental data. Line: fit by the function
$A_{\bm k} \cos(2\omega_{\bm k} t +\phi_{\bm k}) + C_{\bm k}$. 
Here, $(k_x,k_y) = (2\pi/L)(n_x,n_y)$ where $n_{x(y)} = -L/2+1,..., L/2$.
In $\bf A,B$, we set $\phi_{\bm k}=0$, except for $(n_x,n_y)=(3,2),(4,0),(4,1)$.
For $(n_x,n_y) = (3, 2), (4,0),(4,1)$, the fitting curves in $\bf B$ are not shown as the standard deviations of 
the fitted  $\hbar\omega_k/J$ are very large.} 
\label{f.fits_FM_tVMC_vs_exp}
\end{figure*}
\end{center}

\begin{center}
\begin{figure*}
\includegraphics[width=\textwidth]{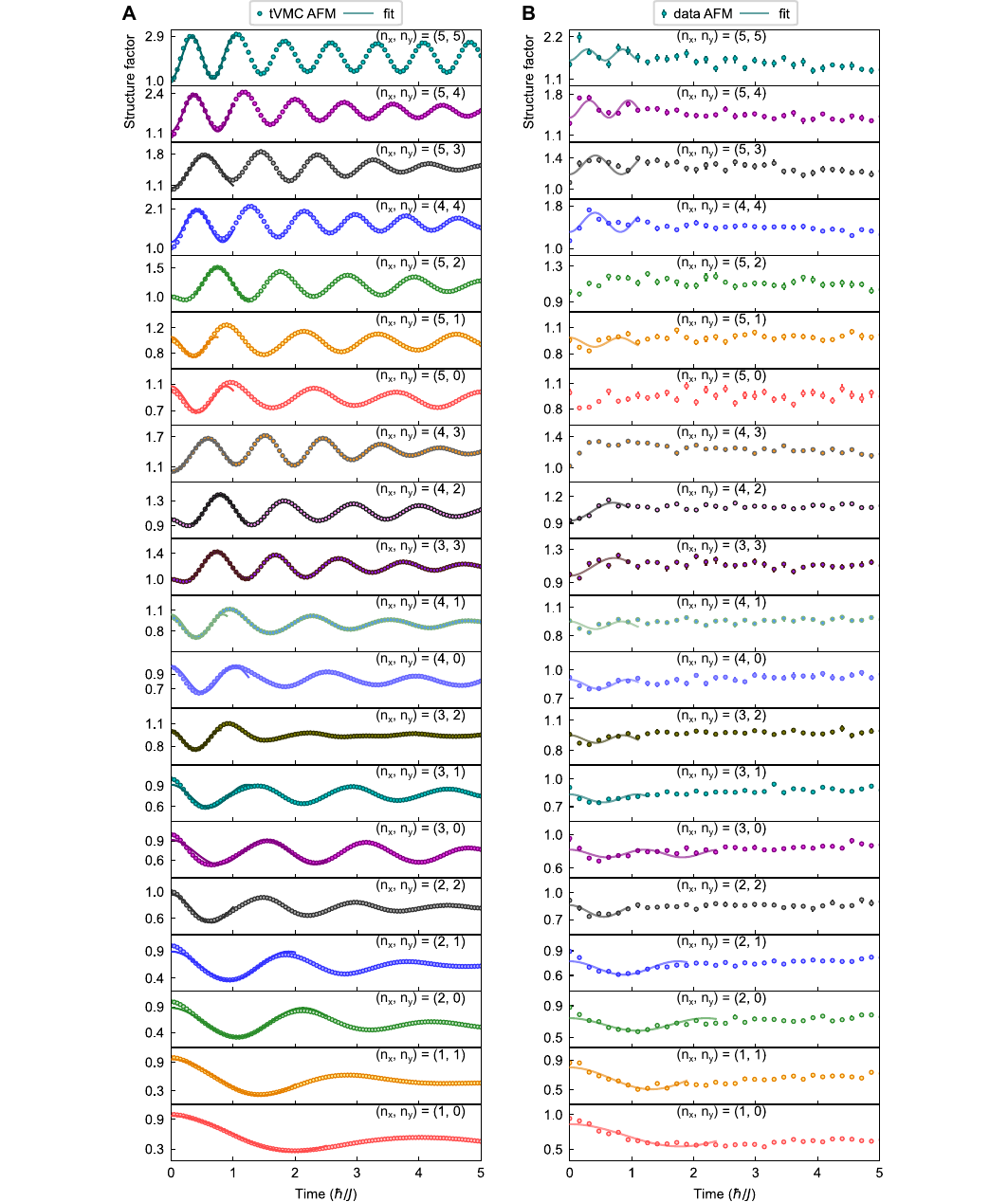}
\caption{\textbf{Full data sets for the time-dependent antiferromagnetic structure factor $S(k_x,k_y;t)$}. 
(Left): tVMC calculations. (Right): experimental data. Line: fit by the function
$A_{\bm k} \cos(2\omega_{\bm k} t +\phi_{\bm k}) + C_{\bm k}$.
Here, $(k_x,k_y) = (2\pi/L)(n_x,n_y)$ where $n_{x(y)} = -L/2+1,..., L/2$.
In $\bf A$, we set $\phi_{\bm k}=0$, except for $(n_x,n_y)=(3,3),(4,2),(5,2)$.
For $(n_x,n_y) = (4, 3), (5,0),(5,2)$, the fitting curves in $\bf B$ are not shown as the standard deviations of 
the fitted  $\hbar\omega_k/J$ are very large.}
\label{f.fits_AFM_tVMC_vs_exp}
\end{figure*}
\end{center}

Figures \ref{f.fits_FM_tVMC_vs_exp} (FM case) and \ref{f.fits_AFM_tVMC_vs_exp} (AFM)
present full data sets for the time-dependent structure factor $S(k_x,k_y;t)$ from the experimental 
data and the data obtained by tVMC calculations. Both experimental and numerical data are fitted with the
same function $A_{\bm k} \cos(2\omega_{\bm k} t +\phi_{\bm k}) + C_{\bm k}$ during the early dynamics.

\section{Principle of the quench spectroscopy method}\label{s.QS}

We  discuss here the bases of quench spectroscopy (QS) introduced in several works
\cite{Menu2018, Frerot2018, Villa2019, Villa2020, Menu2023}, 
as applied to lattice spin models with \emph{periodic} boundary conditions.
The quantity of interest is the time-dependent structure factor 
\begin{eqnarray}\label{Eq:tSF}
S^{\mu\mu}(\bm k, t) & = & 
\frac{1}{N} \sum_{ij} e^{-i \bm k \cdot (\bm r_ i - \bm r_j)} \langle \Psi(t) | \sigma^\mu_i \sigma^\mu_j  |\Psi(t) \rangle \nonumber \\
& = & \langle \Psi(t)| \sigma_{\bm k}^\mu \sigma_{-\bm k}^\mu |\Psi(t)\rangle
\end{eqnarray}
where $\mu = x, y, z$. We have introduced the Fourier transform of the spin operators
\begin{equation}
\sigma_{\bm k}^\mu = \frac{1}{\sqrt{N}} \sum_i e^{-i \bm k \cdot \bm r_i} \sigma_i^\mu~.
\end{equation}
We consider quench experiments which start from a product state 
$|\Psi(0)\rangle = \otimes_{i=1}^N |\psi_i\rangle$, corresponding to the mean-field 
approximation to the ground state of the spin Hamiltonian $H$ of interest. 
If such a ground state has spins forming a
pattern in the $xy$ plane (as is the case in our work), 
elementary excitations (with integer spin $s=1$) are spin 
flips with respect to these spin orientations: They are generated locally by the 
$\sigma_i^z = (\sigma_i^+ + \sigma_i^-)/2$ operator -- $\sigma_i^\pm$ 
being the raising/lowering operators of spin $i$ with respect to the  
$y$ quantization axis, along which the spins are aligned in the initial state. 
Elementary excitations in a periodic system are 
moreover labeled by their momentum, so that delocalized spin flips with 
wavevector $\bm k$ should be generated by the $\sigma^z_{\bm k}$ operator.  

The time-dependent structure factor $S^{zz}(\bm k,t) = S(\bm k, t)$ can  then be rewritten as 
\begin{equation}
S(\bm k, t) = \sum_{nm} e^{i \omega_{nm} t} \langle \Psi(0)| n \rangle \langle m | \Psi(0)\rangle 
\langle n | \sigma_{\bm k}^z\sigma_{-\bm k}^z | m\rangle   
\label{e.Skt}
\end{equation}
where we have introduced the Hamiltonian eigenstates 
$H|n\rangle  = \hbar \omega_n|n\rangle$, and $\omega_{nm} = \omega_n - \omega_m$. 
The time-dependent structure factor therefore oscillates at frequencies $\omega_{nm}$ 
corresponding to transitions between Hamiltonian eigenstates $|m\rangle \to |n\rangle$ 
which are connected by \emph{two} delocalized spin flips at opposite wavevectors $\pm \bm k$, 
generated by the operators $\sigma_{\pm \bm k}^z$. 
In turn, these two states must be both contained in the initial state $|\Psi(0)\rangle$, 
as the transition between them is weighted by the overlaps $\langle \Psi(0)| n \rangle$
and $\langle m | \Psi(0)\rangle$. This means that, if the initial state is chosen so as to 
overlap with the low-lying part of the spectrum of the Hamiltonian, only two-spin-flip 
transitions between low-energy states will contribute to the time dependence of the structure factor. 
In particular, if the transitions $|0\rangle \to |n\rangle$ dominate in the sum Eq.~\eqref{e.Skt} 
(where $|0\rangle$ is the Hamiltonian ground state), the time dependence reveals primarily 
the spectrum of two-spin-flip elementary excitations. More generally, if spin-flip excitations 
correspond to free quasiparticles, their corresponding transition energy remains the 
same regardless of the $|m\rangle \to |n\rangle$ transition, so that the spectrum of 
elementary excitations is revealed regardless of the degree of overlap 
of $|\Psi(0)\rangle$ with the Hamiltonian ground state. 
More precisely, as
\begin{align}
& \langle n | \sigma_{\bm k}^z \sigma_{-\bm k}^z | m\rangle  \nonumber \\
& = \langle n | \left ( \sigma_{\bm k}^+ \sigma_{-\bm k}^- +  \sigma_{\bm k}^- \sigma_{-\bm k}^+ 
+ \sigma_{\bm k}^+ \sigma_{-\bm k}^+ + \sigma_{\bm k}^- \sigma_{-\bm k}^- \right ) |m\rangle
\end{align}
the transition $|m\rangle \to |n\rangle$ is generated by two spin flips 
at opposite wavevectors.
In the case of spin flips corresponding to free quasi-particle excitations 
with a well-defined dispersion relation $\omega_{\bm k}$, 
$S({\bm k}, t)$ oscillates  therefore at frequencies
$\omega_{\bm k} \pm \omega_{-\bm k}$. 
Hence, if $\omega_{\bm k} = \omega_{-\bm k}$ (time-reversal invariance, valid in our case), 
$S({\bm k}, t)$ oscillates at frequencies $\omega_{nm}  = 0,  2 \omega_{\bm k}$.  
This is indeed the prediction of linear spin-wave theory, as discussed in the next section. 

\section{Linear spin-wave theory for XY magnets -- open vs. periodic boundary conditions}\label{s.LSW}

In this section, we review the essential features of spin-wave theory of the XY magnets,
described in many references \cite{Auerbach1994,Mattis-book,Peter2012,Frerot2018}, 
and their relation to quench spectroscopy. 
 
The Hamiltonian for the spin-1/2 XY model with power-law interactions is
\begin{equation}
H  =  - \frac{1}{2} \sum_{i<j} J_{ij}  \left ( \sigma_i^x \sigma_j^x + \sigma_i^y \sigma_j^y \right )\ . 
\label{e.Hdip}
\end{equation} 
We consider the mean-field ground state of this Hamiltonian to always be the 
coherent spin state (CSS) $|{\rm CSS}\rangle = |\rightarrow_y \rangle^{{\otimes}^N}$. 
This is true for the dipolar ferromagnet with couplings $J^{\rm (FM)}_{ij} \geq 0$, 
and, as explained in the main text,  
it becomes true for antiferromagnetic interactions on a bipartite lattice 
(such as the square lattice) when rotating around the $z$ axis 
by $\pi$ the spins on the A sublattice, 
so that  $J^{\rm (AFM)}_{ij} = (-1)^{i+j+1} J^{\rm (FM)}_{ij}$. 

Spin-wave theory for the above models is built by mapping spins onto bosons 
via the Holstein-Primakoff (HP) transformation
$\sigma_i^y  =   1 - 2 n_i$,
$\sigma_i^z  =  \left ( \sigma_i^+ + \sigma_i^- \right) = \left ( \sqrt{1-n_i} ~b_i + b_i^\dagger\sqrt{1-n_i} \right ) \nonumber$, 
$\sigma_i^x  =  -i\left ( \sigma_i^+ - \sigma_i^- \right) = -i \left ( \sqrt{1-n_i} ~b_i  - b_i^\dagger\sqrt{1-n_i}  \right ) \nonumber$. 
Here $b_i$, $b_i^\dagger$ are bosonic operators, and $n_i = b_i^\dagger b_i$. 
Replacing the spin operators with HP bosons, and truncating the Hamiltonian to quadratic order, one obtains
$H_{XY} \approx  E_{\rm CSS} + H_2$  where 
$E_{\rm CSS} = -\sum_{i<j} J_{ij}/2$ 
is the energy of the CSS, and
\begin{equation}
H_2 = \sum_{i<j} J_{ij}  \Big [ (n_i + n_j) - \frac{1}{2} 
\left ( b_i b_j + b^\dagger_i b^\dagger_j  + b_i^\dagger b_j + b^\dagger_j b_i   \right ) \Big ] 
\label{e.H2real}
\end{equation}
is the Hamiltonian describing quadratic fluctuations around the mean-field solution. 
It can be Bogolyubov diagonalized 
via a linear transformation on the bosonic operators. 
We discuss in the next section the cases of periodic vs. open boundary conditions, 
as both are relevant for this work. 

\subsection{Periodic boundary conditions}\label{s.PBC}

\begin{figure*}
\centering
\includegraphics[width=0.5\linewidth]{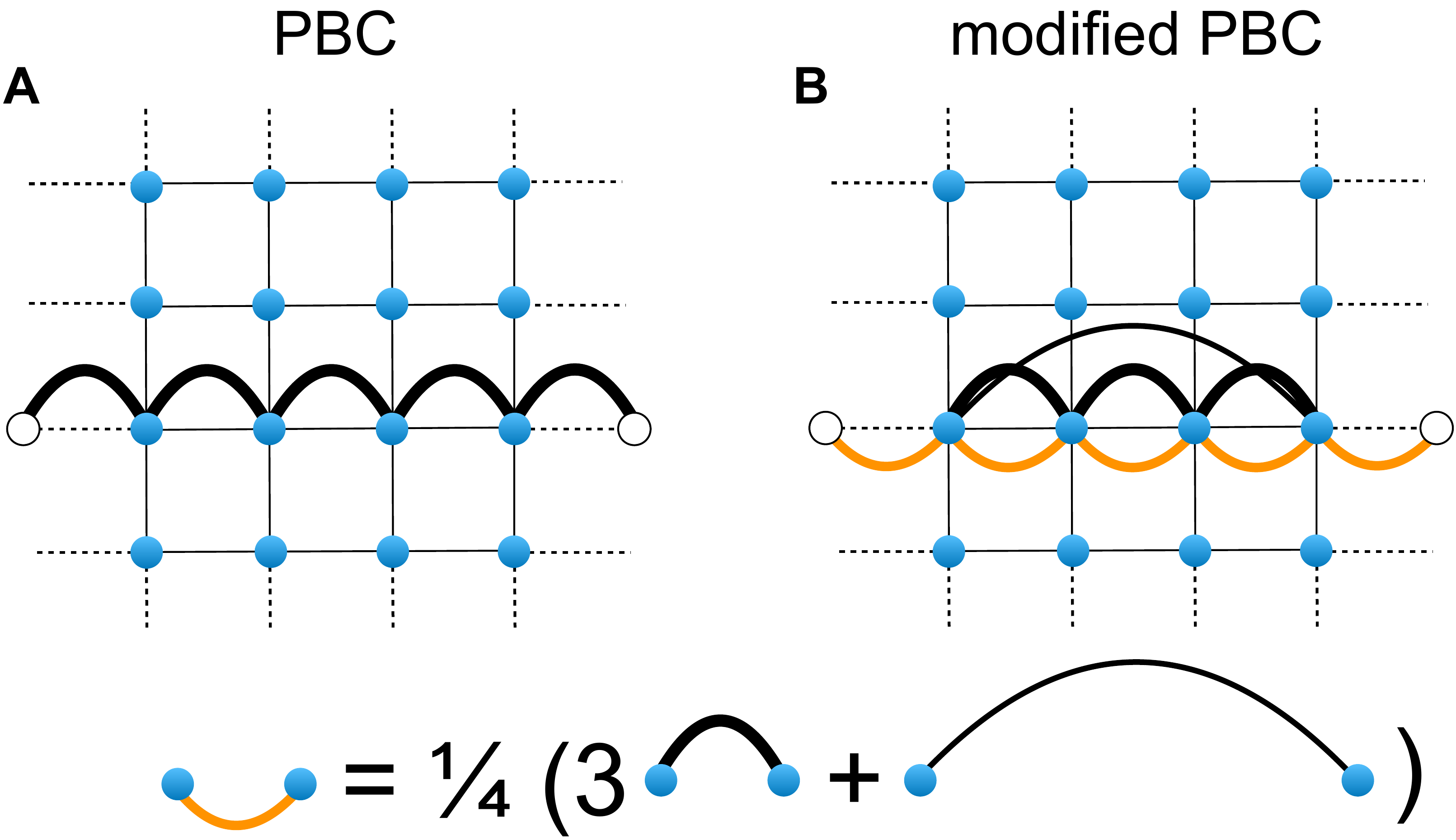}
\caption{{\bf Periodic boundary conditions (PBC) vs. modified periodic boundary conditions (mPBC)}. 
Here we exemplify how a specific form of couplings -- those connecting nearest neighbors -- 
change when going from PBC to mPBC. 
\textbf{(A)} The black arcs indicate nearest-neighbor couplings for PBC. 
On a torus, these couplings connect boundary sites to the periodic image of the sites 
on the opposite boundary (indicated with the open circles); 
\textbf{(B)} the orange arcs indicate the same couplings for mPBC, and are given by the average 
between the nearest-neighbor couplings and the dipolar coupling between opposite sites, 
which are nearest neighbors on the torus (see the formula indicated in the figure).}
\label{f.mBPC}
\end{figure*}

When using periodic boundary conditions, the couplings $J_{ij}$ 
are calculated by taking distances between sites on a torus, 
with a maximal distance of $L/2$ for a $L\times L$ square lattice: 
in the case of dipolar interactions  $J^{\rm (FM)}_{ij,\rm PBC} =  J/(r^{(t)}_{ij})^3$, 
with a distance on the torus 
$r^{(t)}_{ij} = [\min^2(|x_i-x_j|,L-|x_i-x_j|) + \min^2(|y_i-y_j|,L-|y_i-y_j|)]^{1/2}$ (see Fig.~\ref{f.mBPC}A), 
and as before $J_{ij, \rm PBC}^{\rm (AFM)} = (-1)^{i+j+1} J_{ij,\rm PBC}^{\rm (FM)}$. 

To perform the Bogolyubov diagonalization of the quadratic bosonic Hamiltonian  
on a periodic lattice, we consider the Fourier transformed Bose operators,
$b_{\bm k} = \sum_i e^{-i \bm k \cdot \bm r_i} b_i/\sqrt{N}$.
Introducing the operators $a_{\bm k}$ and $a^\dagger_{\bm k}$ such that
$a_{\bm k} = u_{\bm k} b_{\bm k} + v_{\bm k} b^\dagger_{-\bm k}$ leads to the diagonal form 
$H_2 = \sum_{\bm k} \hbar \omega_{\bm k} a_{\bm k}^\dagger a_{\bm k} +  \sum_{\bm k} (\epsilon_{\bm k} -  {\cal A}_{\bm k})/2$
where ${\cal A}_{\bm k}  =   J (\gamma^{(t)}_0 - \gamma^{(t)}_{\bm k}/2)$, and $\omega_{\bm k}$ is given in the main text.
Here (and contrarily to the main text) $\gamma^{(t)}_{\bm k}$ is the Fourier transform of the dipolar interactions 
{\it defined on a torus}, 
$\gamma^{(t)}_{\bm k} = N^{-1} \sum_{i\neq j} \eta^{i+j+1}/(r^{(t)}_{ij})^3$ with $\eta =1 (-1)$ for the FM (AFM). 
The bosonic quasiparticles associated with the Bogolyubov operators 
$a_{\bm q}, a_{\bm q}^\dagger$ are called \emph{magnons}, and they represent the linearized excitations of the system.  

\begin{figure*}
\centering
\includegraphics[width=\textwidth]{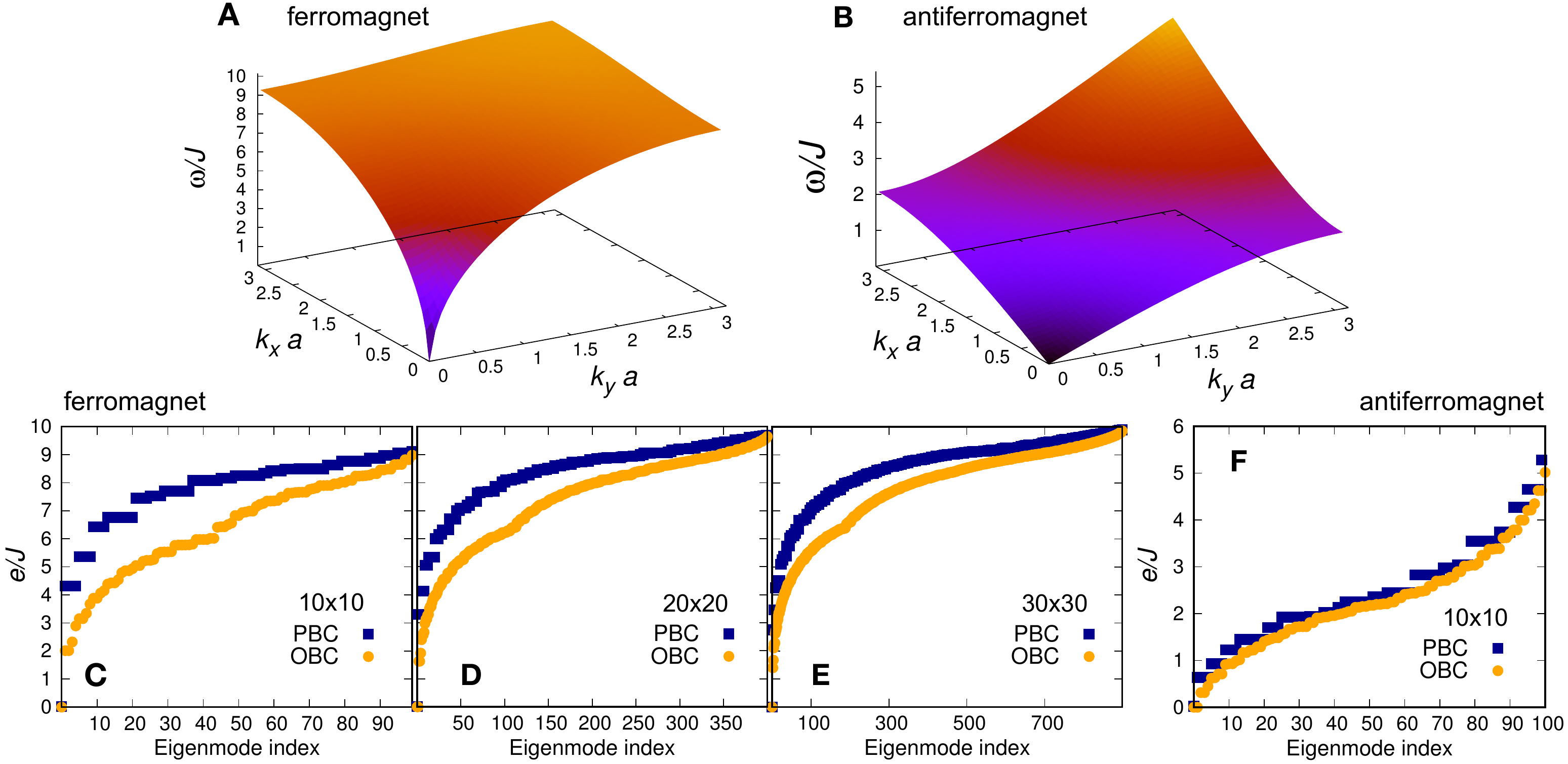}
\caption{{\bf Dispersion relation of the dipolar 2d XY model from linear spin-wave theory.} 
\textbf{(A)}: ferromagnet; \textbf{(B)}: antiferromagnet.
\textbf{(C-E)}: Eigenenergies from linear spin-wave theory for the 2d XY dipolar ferromagnet. 
We contrast periodic (PBC) with open (OBC) boundary conditions, for three system sizes: 
\textbf{(C)}: $10\times 10$; \textbf{(D)}: $20\times 20$; \textbf{(E)}: $30\times 30$.
\textbf{(F)}: Eigenenergies from linear spin-wave theory for the 2d XY dipolar antiferromagnet, 
contrasting periodic (PBC) with open (OBC) boundary conditions on a $10\times 10$ lattice.}
\label{f.LSW_dispersion}
\end{figure*}

The dispersion relations of the dipolar ferromagnet and antiferromagnet  are shown in 
Fig.~\ref{f.LSW_dispersion}A,B. They exhibit the characteristic nonlinear behavior 
$\omega_{\bm k} \sim \sqrt{k}$  at small $k$ for the ferromagnet, as well as linear 
behavior  $\omega_{\bm k} \sim k$ for the antiferromagnet. 

The time-dependent structure factor has a simple form within linear spin-wave theory \cite{Frerot2018}:
\begin{eqnarray}
S_{\rm LSW}({\bm k},t) &=&  \langle b^\dagger_{\bm k} b^\dagger_{\bm k} +  b_{-\bm k} b^\dagger_{-\bm k} - b_{\bm k} b_{-\bm k}  - 
b^\dagger_{\bm k} b^\dagger_{-\bm k} \rangle   \nonumber \\
&=& C_{\bm k} + A_{\bm k} \cos(2\omega_{\bm k} t)\ ,
\label{e.TSF_LSW}
\end{eqnarray}
where 
$C_{\bm k} = 1- \gamma^{(t)}_{\bm k}/(2\gamma^{(t)}_0)$ and $A_{\bm k}= \gamma^{(t)}_{\bm k}/(2\gamma^{(t)}_0)$.
It features a term at zero frequency (corresponding to the constant term) 
as well as a term at frequency $2\omega_{\bm k}$, 
as expected from the discussion of quench spectroscopy in Sec.~\ref{s.QS}. 
Moreover,  LSW theory yields a relation between the constant 
$C_{\bm k}$ and the frequency $\omega_{\bm k} = J\gamma^{(t)}_0 \sqrt{2C_{\bm k}-1}~$.  
A similar relationship holds for the oscillation amplitude $A_{\bm k}$: 
\begin{equation}
\omega_{\bm k} = J\gamma^{(t)}_0 \sqrt{1-2A_{\bm k}}~.
\label{e.omegakAk}
\end{equation}

\subsection{Open boundary conditions} 

For lattices with open boundary conditions, one can still diagonalize the 
Hamiltonian of Eq.~\eqref{e.H2real} via a generalized Bogolyubov transformation
$a_\alpha = \sum_i ( u_i^{(\alpha)} b_i + v_i^{(\alpha)} b_i^\dagger)$
which brings the quadratic Hamiltonian to the form 
$H_2 = \sum_\alpha \hbar \omega_\alpha a_{\alpha}^\dagger a_{\alpha} + {\rm const} $. 
To do so, one builds the $N \times N$ matrices $A$ and $B$, with 
$2A_{ij} = \left (\sum_k J_{ik} ~\delta_{ij} -  J_{ij}/2 \right ) $ and $B_{ij} = J_{ij}/4$. 
One then diagonalizes the $2N \times 2N$ non-Hermitian matrix  
$\begin{pmatrix} A &  B \\ - B & -A \end{pmatrix}$ with a matrix of right 
eigenvectors $\begin{pmatrix} U &  -V \\ -V & U \end{pmatrix}$  where 
$U_{i\alpha} = u_i^{(\alpha)}$ and $V_{i\alpha} = v_i^{(\alpha)}$ are $N \times N$ 
matrices \cite{blaizot-ripka,FrerotR2015}, assumed here to be real-valued 
(as is the case for the models studied in this work).

\subsection{Effect of boundary conditions on the linear spin-wave eigenspectrum}\label{s.LSW.boundary}

The experiments of this work are performed on lattices with open boundary conditions (OBC). 
It is thus important to compare the linear spin-wave spectrum on open lattices 
with that of lattices with periodic boundary conditions (PBC). 
Figures.~\ref{f.LSW_dispersion}C,F show the ordered eigenfrequencies spectrum 
$\omega_\alpha$ for square lattices with PBC and OBC for both the XY dipolar 
ferromagnet and the antiferromagnet.  
The eigenfrequencies of the ferromagnet with OBC are systematically smaller 
than those of the same model with PBC: this is a consequence of the dipolar interactions, 
which enhance the role of boundary conditions with respect to \emph{e.g.} nearest-neighbor interactions. 
In particular the discrepancy between the OBC and PBC spectra persists for large lattices, 
and it is still visible when tripling the linear size of the lattice (from $10\times 10$ to $30\times 30$). 
Contrarily, the antiferromagnet is less sensitive to boundary conditions, 
and already on the $10\times 10$ lattice the OBC and PBC spectra are very close. 
This is a manifestation of the effectively short-ranged nature of the antiferromagnetic interactions, 
due to frustration: sites close to the boundary of an OBC system and sites in the bulk 
experience almost the same interactions, and boundary effects are less important. 

The results in Fig.~\ref{f.LSW_dispersion}C,E may suggest that quench spectroscopy experiments 
on the ferromagnet on \emph{e.g.} a $10\times 10$ lattice will reconstruct a very different dispersion relation 
than the one predicted for the PBC system. Moreover, this reconstruction will be complicated 
by the fact that the eigenmodes of the ferromagnet do not correspond to plane waves, 
and therefore the momentum-dependent spin-flip operators $\sigma_{\pm \bm k}^z$ 
excite a superposition of them. 
This difficulty is in fact overcome when considering the system dynamics at sufficiently short times,  
thanks to a short-time projective equivalence between the OBC system and a system with modified PBC
as we explain in the next section. 

\subsection{Correcting spin-wave theory for the single-site constraint on correlations} 
\label{s.corr}

\begin{figure*}
\centering
\includegraphics[width=0.5\textwidth]{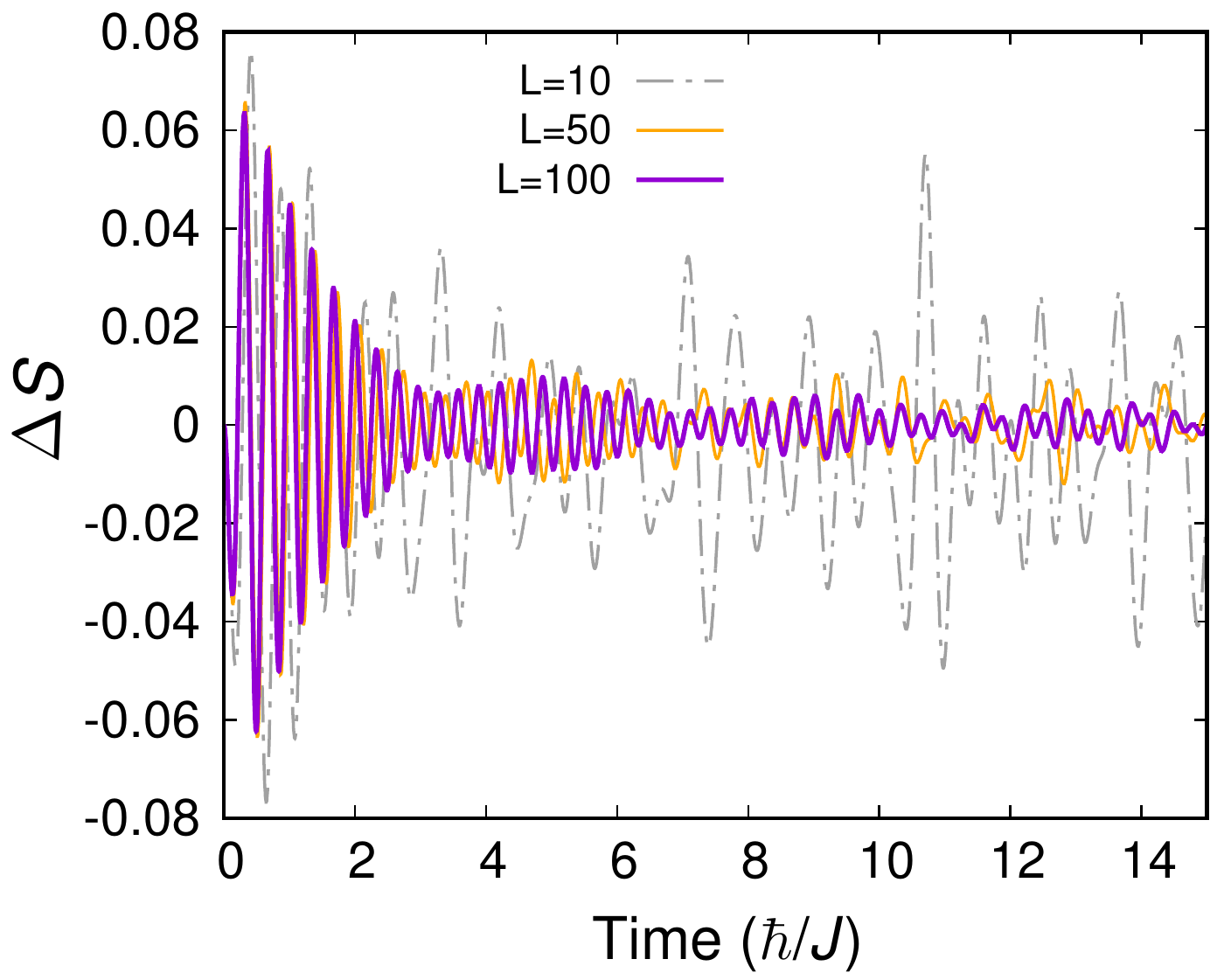}
\caption{{\bf Non-linear correction to LSW theory, accounting for the constraint $(\sigma^z)^2 = \bf 1$.} 
We plot the correction $\Delta S (t)$ (Eq.~\eqref{e.DeltaS}) for the FM with mPBC, 
for three different sizes $N = L \times L$. The correction is seen to scale to zero with system size due to dephasing.}
\label{f.DeltaS}
\end{figure*}

A central aspect of spin-$1/2$ systems missed by linear spin-wave theory is the fundamental 
property that $(\sigma^z_i)^2=\mathbf{1}$. This is a strong, local constraint which 
cannot be reproduced by the linearized Holstein-Primakoff transformation, giving instead 
$(\sigma^z_i)^2 \approx \mathbf{1} + 2 b_i^\dagger b_i + b_i^2 + (b_i^\dagger)^2$.
Nonetheless the constraint can be imposed by hand on real-space correlations, 
and then propagated onto their Fourier transform.
Given that 
\begin{equation}
\frac{1}{N} \sum_{\bm k} S(\bm k,t) = \frac{1}{N} \sum_{i} \langle (\sigma_i^z)^2 \rangle
\end{equation}
the condition $(\sigma_i^z)^2=\bf 1$ amounts to requesting that 
\begin{equation}
\frac{1}{N} \sum_{\bm k} S(\bm k,t) = 1~.
\end{equation}
To enforce this sum rule, the structure factor obtained via standard spin-wave theory, Eq.~\eqref{e.TSF_LSW}, can be modified to
$S({\bm k},t) = S_{\rm LSW}(\bm k,t) + \Delta S(t)$, where
\begin{equation}
\Delta S(t) =  1- \frac{1}{N} \sum_{\bm k} S_{\rm LSW}(\bm k,t) = -\frac{1}{N} \sum_{\bm k} A_{\bm k} \cos(2\omega_{\bm k}t)
\label{e.DeltaS}
\end{equation} 
using the fact that $\frac{1}{N} \sum_{\bm k} C_{\bm k} = 1$ since $\sum_{\bm k} \gamma_{\bm k} = 0$ 
(no self-interaction between spins).  This correction 
has a very important consequence for quench spectroscopy. 
When  $\Delta S$ is comparable to the amplitude $A_{\bm k}$ of the oscillations of 
the time-dependent structure factor, $|\Delta S(t)| \gtrsim |A_{\bm k}|$, the spectral content of the 
oscillations can be significantly altered compared to linear spin-wave theory, especially at short times. 
As we will discuss in more details in Sec.~\ref{s.omega_shift}, the correction leads to a significant 
\emph{shift} of the frequency estimated from the short-time oscillations of the time-dependent structure factor, 
compared with the linear spin-wave prediction. 

On the contrary, the $\Delta S$ correction alters minimally the offset $C_{\bm k}$ of the oscillations. 
Indeed, averaging over time gives
\begin{equation} 
\overline{\sum_{\bm k} S_{\rm LSW}(\bm k,t)} = N + A_{\bm 0},
\end{equation}
and, given that  $A_{\bm 0} = 1/2$, one obtains that 
$\overline{\Delta S(t)} = - \frac{1}{2N}$, which becomes negligible for large $N$.  
The amplitude $A_{\bm k}$ of the oscillations of the time-dependent structure factor 
should be instead affected by the correction. Nonetheless, as we shall see later, 
in spite of this correction it is still possible to extract reliably the dispersion relation
of linear spin waves from the amplitude via Eq.~\eqref{e.omegakAk}.

A fundamental remark is in order concerning the scaling of the $\Delta S$ correction 
with system size, and its short-time vs. long-time behavior. 
From its definition (Eq.~\eqref{e.DeltaS}), 
$\Delta S$ is a sum of oscillating terms at different frequencies $2\omega_{\bm k}$. 
Hence, since these frequencies are generically incommensurate to each other, 
at sufficiently long times the sum will experience \emph{dephasing}, 
namely $\sum_{\bm k} A_{\bm k} \cos(2\omega_{\bm k}t) \sim {\cal O}(\sqrt{N})$ as for a sum of 
$N-1$ incoherent terms. 
As a consequence we expect that, for a time $t > \pi/\omega_{\rm min}$, 
where $\omega_{\rm min} = \min_{\bm k \neq 0} \omega_{\bm k}$, $\Delta S(t)$ becomes of order $1/\sqrt{N}$. 
This is confirmed in Fig.~\ref{f.DeltaS}, 
which shows the explicit time dependence of $\Delta S(t)$ for various system sizes in the FM case. 
This implies that the $\Delta S$ correction is fundamentally a \emph{finite-size} and \emph{finite-time} correction, 
disappearing at long times (but not at short time) in the thermodynamic limit. Standard LSW theory is therefore expected to 
emerge as the correct quantitative description of systems such as the 2d FM dipolar XY model only 
for large system sizes and at long times. 
These sizes, and long enough coherence times, should become accessible to Rydberg-atom simulators in the forthcoming future.

\section{Projective equivalence between OBC and modified-PBC lattices at short times} \label{s.short_times}

\subsection{Momentum-sector decomposition of the Hilbert space and of the Hamiltonian}

As already pointed out above, a PBC lattice has the topology of a torus, 
and its properties are invariant under translations $T_{\bm d}$  
by a vector $\bm d = (d_x,d_y)$ along the torus.  
The eigenvectors of the Hamiltonian with PBC are therefore also eigenvectors 
of the symmetry operator $T_{\bm d}$,
namely states $|\Psi_{\bm Q}\rangle$ with definite total momentum ${\bm Q}$, such that 
$T_{\bm d} |\Psi_{\bm Q}\rangle = e^{i\bm Q \cdot \bm d} |\Psi_{\bm Q}\rangle$. 

An OBC lattice can also be thought of as being wrapped to form a torus. 
However, the couplings $J_{ij}$ depend on the standard euclidean distance between sites 
(and not on the distance on the torus), $J_{ij} = J/r^3_{ij}$ 
with $r_{ij} = [(x_i - x_j)^2 + (y_i - y_j)^2]^{1/2}$. 
As a consequence the system with OBC is not invariant under translations $T_{\bm d}$  on the torus. 

Yet, even for the OBC lattice, it is useful to keep a picture of Hilbert space as being 
structured into different momentum sectors, with related projectors 
${\cal P}_{\bm Q} = \sum_{\Psi_{\bm Q}} |\Psi_{\bm Q}\rangle \langle \Psi_{\bm Q}|$, 
such that $\sum_{\bm Q} {\cal P}_{\bm Q} = \mathds{1}$. The Hamiltonian $H_{\rm OBC}$ on a OBC lattice, 
when written on the basis of momentum eigenstates, 
will then possess a diagonal part as well as an off-diagonal one $H_{\rm OBC} = H_{\rm D} + H_{\rm OD}$ with
\begin{equation}
H_{\rm D} = \sum_{\bm Q} {\cal P}_{\bm Q} H_{\rm OBC}  {\cal P}_{\bm Q}\ , \ 
H_{\rm OD} = \sum_{\bm Q \neq \bm Q' } {\cal P}_{\bm Q} H_{\rm OBC}  {\cal P}_{\bm Q'}.
\end{equation} 
In the case of the Hamiltonian Eq.~\eqref{e.Hdip}, the diagonal and off-diagonal parts are readily identifiable  as
\begin{eqnarray}
H_{\rm D} & = &  \frac{J}{2} \sum_{\bm k} {\gamma}_{\bm k} \left ( \sigma_{\bm k}^x \sigma_{-\bm k}^x + 
\sigma_{\bm k}^y \sigma_{-\bm k}^y \right )   \nonumber \\
H_{\rm OD} & = &  \frac{J}{2}  \sum_{\bm k \neq \bm k'} \Gamma_{\bm k, \bm k'} 
\left (\sigma_{\bm k}^x \sigma_{-\bm k'}^x + \sigma_{\bm k}^y \sigma_{-\bm k'}^y \right ) 
\label{e.HOBC}
\end{eqnarray}
where 
\begin{equation}
 J \Gamma_{\bm k, \bm k'} = \frac{1}{N} \sum_{ij} e^{i({\bm k} \cdot \bm r_i - {\bm k'} \cdot \bm r_j)} ~J_{ij}
\end{equation}
and $\gamma_{\bm k} = \Gamma_{\bm k, \bm k}$ -- coinciding with the one used in the main text. 
The diagonal part $H_D$ is an effective Hamiltonian with periodic boundary conditions, 
and modified couplings $\tilde J_{ij, \rm mPBC} = \tilde J({\bm r^{(t)}_{ij}})$ 
where ${\bm r}^{(t)}_{ij}$ is a distance on the torus, and 
\begin{equation}
\tilde J({\bm r}) = \frac{J}{N} \sum_{\bm k} e^{-i\bm k \cdot \bm r} \gamma_{\bm k} 
= \frac{1}{N} \sum_{hl} J_{hl,\rm OBC} ~ \delta_{{\bm r}^{(t)}_{hl}, \bm r}  \ .
\label{e.tildeJ}
\end{equation}
Thus, $\tilde J({\bm r})$ is the average coupling between all sites on the OBC lattice 
that possess that same distance ${\bm r}^{(t)}_{hl}={\bm r}$ on the torus. 
In the following we denote the Hamiltonian $H_{\rm D}$ as the Hamiltonian with modified PBC (mPBC). 

All the results of spin-wave theory for PBCs, described in Sec.~\ref{s.PBC}, 
apply as well to the Hamiltonian with mPBC after the simple substitution 
$\gamma^{(t)}_{\bm k} \to \gamma_{\bm k}$. 

\begin{figure*}
\centering
\includegraphics[width=0.8\textwidth]{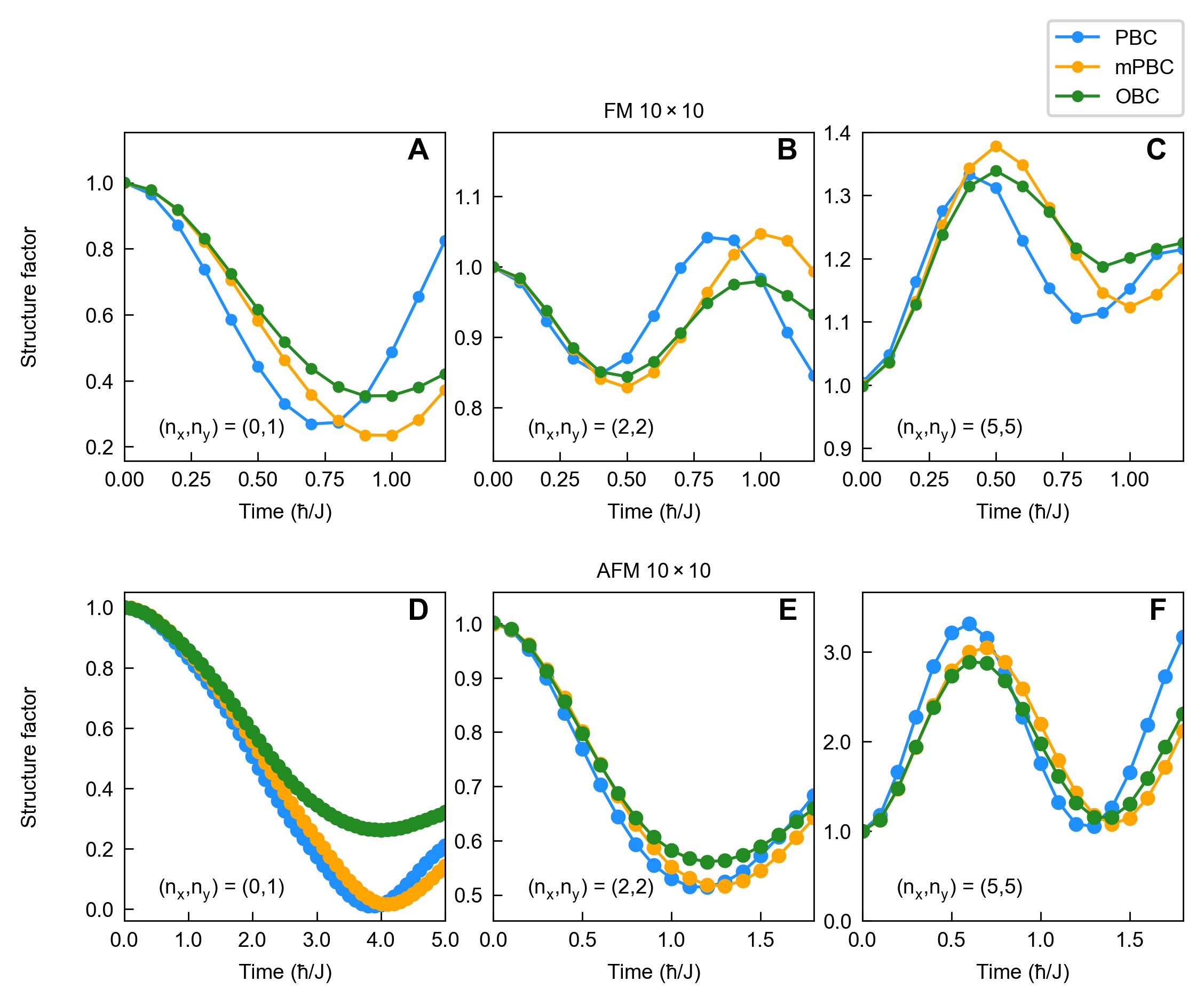}
\caption{{\bf Time-dependent structure factor at short times for different boundary conditions}, obtained by 
tVMC on a $10\times 10$ lattice. 
Comparison between periodic (PBC), modified-periodic (mBPC) and open (OBC) boundary conditions 
for three wavevectors ${\bm k} a/\pi = (0,1)$, $(2,2)$ and $(5,5)$ 
for the ferromagnet (panels \textbf{(A)}, \textbf{(B)} and \textbf{(C)}) and the antiferromagnet (panels \textbf{(D)}, \textbf{(E)} and \textbf{(F)}).}
\label{f.OBC_modPBC}
\end{figure*}

\subsection{Short-time projective equivalence with uniform initial states}

The momentum structure of the OBC Hamiltonian \eqref{e.HOBC} 
is relevant when considering unitary evolutions that start in a \emph{definite} 
momentum sector. This is the case of the homogeneous initial states considered in 
this study, namely the CSS state, which is a momentum eigenstate with ${\bm Q}=0$. 

As a consequence, inserting the completeness relation on the momentum basis
$\sum_{\bm Q} {\cal P}_{\bm Q} = \mathds{1}$ in Eq.\eqref{Eq:tSF}, 
the time-dependent structure factor of the OBC system can be rewritten as
\begin{equation}
S(\bm k,t) =  S_{\rm mPBC} (\bm k,t) + \delta S (\bm k,t)
\end{equation}
where 
\begin{equation}
S_{\rm mPBC} (\bm k,t) = 
\langle {\rm CSS} | e^{i H_{\rm D} t/\hbar} \sigma_{\bm k}^z \sigma_{-\bm k}^z  e^{-i H_{\rm D} t/\hbar}  |{\rm CSS} \rangle 
\end{equation}  
is related to the dynamics governed by $H_{\rm D}$, restricted to the zero-momentum sector in 
which it is initialized. The term  
\begin{align} 
& \delta S (\bm k,t)  = \nonumber \\
& \sum_{\bm Q\neq 0} \langle {\rm CSS} | {\cal P}_0 
e^{i H_{\rm OD} t/\hbar} {\cal P}_{\bm Q} \sigma_{\bm k}^z \sigma_{-\bm k}^z  {\cal P}_{\bm Q} 
e^{-i H_{\rm OD} t/\hbar} {\cal P}_0 |{\rm CSS} \rangle 
\end{align}
is the contribution related to the leaking of the dynamics into non-zero momentum 
sectors under the effect of $H_{\rm OD}$. 

At the beginning of the dynamics $S(\bm k,0) = S_{\rm mPBC} 
(\bm k,0)$ and  $\delta S (\bm k,0)=0$ by virtue of the choice of the initial state. 
Therefore there must be an interval $[0, t_{\rm PE}]$ in the early evolution during which 
\begin{equation}
t\in [0, t_{\rm PE}]: S(\bm k,t) \approx S_{\rm PBC} (\bm k,t)\ .
\end{equation}
Thus, over this early evolution time, the dynamics of the time-dependent structure 
factor is (nearly) \emph{projectively equivalent} to that of a system with the same lattice 
structure but with PBC, and possessing the modified couplings of Eq.~\eqref{e.tildeJ}.  
The duration $t_{\rm {PE}}$ of the approximate projective equivalence 
(OBC $\approx$ modified PBC) is difficult to estimate a priori, but it certainly 
extends to longer times the larger the system size, as for $N\to \infty$ the three 
lattice geometries -- OBC, PBC and modified PBC -- will eventually give the same physics. 

We have numerically tested the effective duration 
of the projective equivalence by comparing the dynamics of the time-dependent 
structure factor in the OBC system with that of modified-PBC system. 
Fig.~\ref{f.OBC_modPBC} shows the tVMC predictions for the time-dependent 
structure factor of a $10\times 10$ lattice with PBC, OBC and mPBC, 
both for the case of the dipolar FM and AFM, 
and for three representative wavevectors going from the 
center to the edge of the Brillouin zone.  We observe a nearly exact projective equivalence 
between OBC and mPBC dynamics at short times, and 
an approximate projective equivalence persists at times of 
the order of an interaction cycle $J^{-1}$  
-- enough to observe a first complete oscillation of the time-dependent structure factor. 
Over this oscillation the modification of the couplings in the PBC system 
is what it takes to quantitatively account for the shift 
of the oscillation frequency when going from (standard) PBC to OBC. 

We therefore conclude that, when the initial state is an eigenstate 
of the lattice momentum, the short-time dynamics of the time-dependent 
structure factor of an OBC lattice allows for the reconstruction of the excitation 
spectrum of an effective system with PBC -- namely a spectrum which admits the 
wavevector ${\bm k}$ as a meaningful label of the eigenfrequencies, 
and therefore which defines a dispersion relation in the proper sense.  
Yet  the actual short-time data may still
show a significant shift from the prediction of LSW theory, Eq.~\eqref{e.TSF_LSW}, 
due to the local constraint for spin-1/2 systems (Eq.~\eqref{e.DeltaS}), as we discuss in the next section.

\subsection{Effect of the non-linear correction to the time-dependent structure factor} 
\label{s.omega_shift}

\subsubsection{Effect on the short-time dynamics} 

\begin{figure*}
\centering
\includegraphics[width=0.8\textwidth]{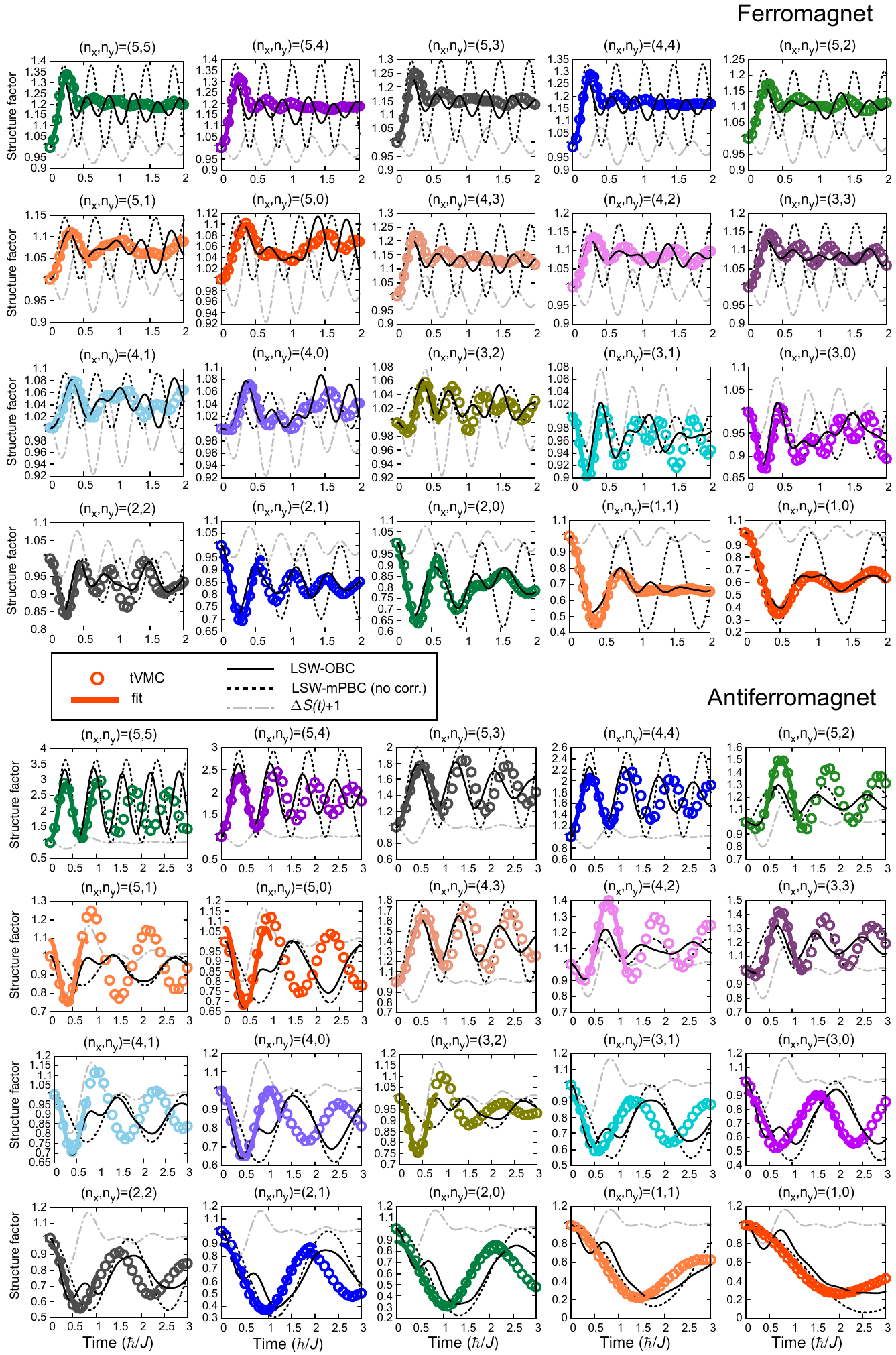}
\caption{{\bf Time-dependent structure factor from tVMC, compared to standard LSW theory and the $\Delta S$ correction}. 
All the data refer to a $10\times 10$ lattice. 
In each window, associated to a different wavevector, we show the tVMC data for a system with OBC; 
their short-time fit to a cosine function; the LSW prediction for the OBC system (including the $\Delta S$  correction); 
the LSW prediction for the mPBC system without the $\Delta S$ correction; 
and the $\Delta S$ correction itself (shifted by $+1$ in order to bring it in the range of the other curves).}
\label{f.FITS_tVMC}
\end{figure*}

\begin{figure*}
\centering
\includegraphics[width=0.7\textwidth]{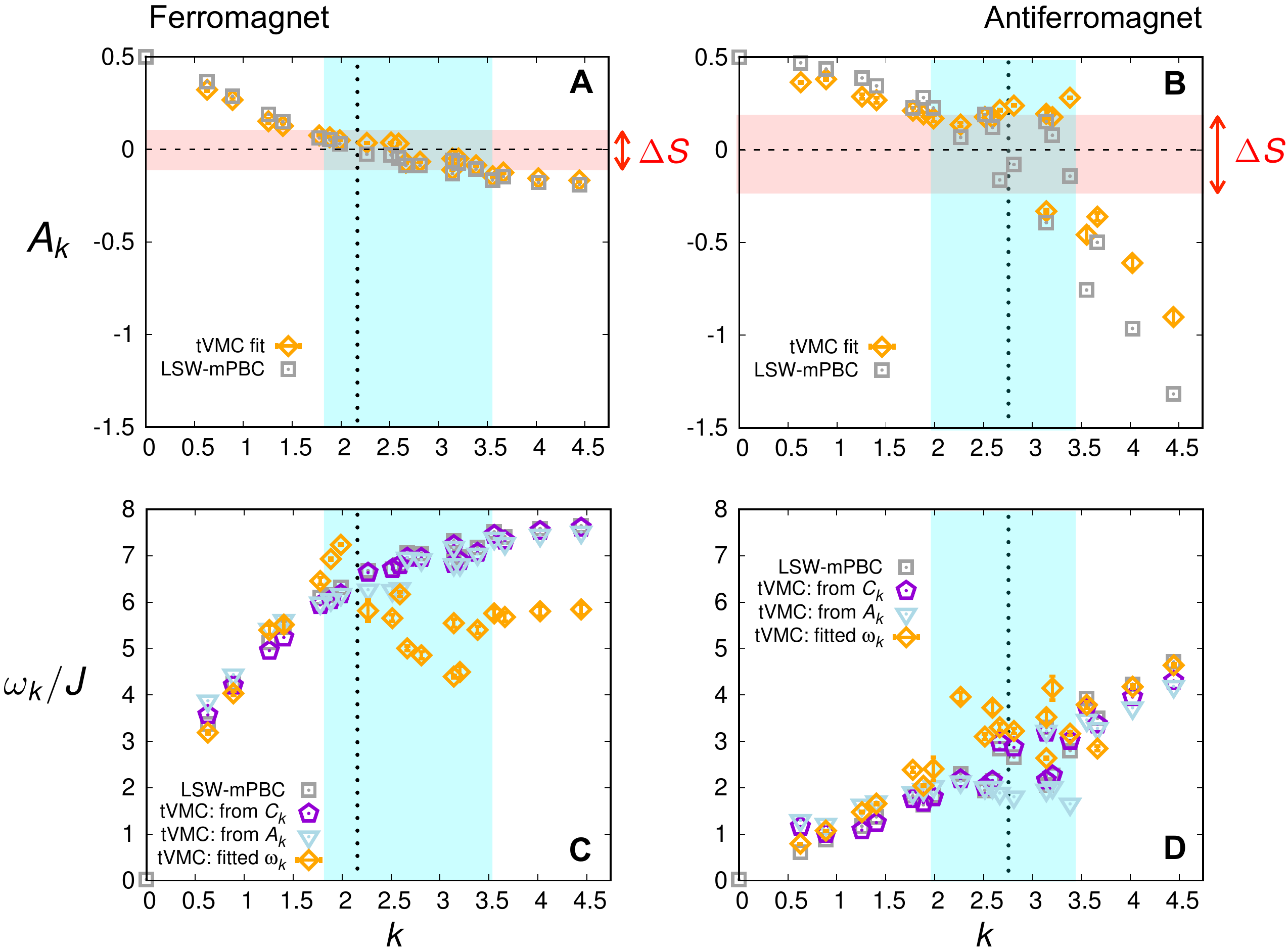}
\caption{{\bf Results from the fits of the tVMC data, compared with the predictions of LSW theory} 
($10\times 10$ array). 
({\bf A}) and ({\bf B}) (FM and AFM respectively) show the $A_{\bm k}$ amplitude from fits to 
tVMC data and from LSW theory on the mPBC system. The red-shaded region shows the amplitude of 
the correction $\Delta S$; the cyan-shaded region marks the interval in wavevector moduli for which 
$|A_{\bm k}|$ is smaller than the amplitude of the oscillations of $\Delta S$;
the horizontal dashed line marks the zero of the ordinates; and the vertical dotted line the wavevector 
modulus at which $A_{\bm k}$ changes sign. Panels ({\bf C}) and ({\bf D}) (for the FM and AFM respectively) 
show the frequency $\omega_{\bm k}$ extracted from the tVMC data by direct fitting, and from the fitted 
$C_{\bm k}$ and $A_{\bm k}$ coefficients, as well as the prediction from LSW theory with mPBCs.}
\label{f.fit_results}
\end{figure*}

As shown in Sec.~\ref{s.corr}, taking into account the single-spin constraint $\langle {\sigma_i^z}^2 \rangle = \mathbf{1}$ 
for finite-size spin-1/2 systems imposes a significant modification to the simple 
spin-wave predictions for the time-dependent structure factor, Eq.~\eqref{e.TSF_LSW}. 
Figure~\ref{f.FITS_tVMC} shows the correction $\Delta S(t)$ of Eq.~\eqref{e.DeltaS} 
as a function of time for the $10\times 10$ FM and AFM, and compares it to the standard 
LSW prediction $S_{\rm LSW}(t)$ for mPBCs at each wavevector. 
One observes that the correction $\Delta S$ can become comparable or even exceed the amplitude 
$A_{\bm k}$ of the fluctuations of $S_{\rm LSW}({\bm k},t)$, for both the FM and the AFM. 
This is because, for both models, the amplitude $A_{\bm k}$ \emph{changes sign} 
(from positive to negative) when going from the center to the edge of the Brillouin zone (BZ), 
as shown in Fig.~\ref{f.fit_results}A,B. Hence the wavevectors $\bm k$ for which $A_{\bm k}$ 
nearly vanishes are strongly affected by the correction. 
In the FM case, the $A_{\bm k}$ amplitudes are actually comparable to the $\Delta S$ 
correction for most wavevectors except for those close to the center of the BZ ($k \to 0$), 
which thus exhibit the smallest fluctuations of the extracted frequencies. 
For the AFM instead, the $A_{\bm k}$ amplitudes can exceed significantly the $\Delta S$ correction, 
both at the center of the BZ as well as towards its edge. 

When $|\Delta S(t)| \gtrsim |A_{\bm k}|$, the time dependence of the structure factor 
can be significantly altered with respect to the standard LSW prediction -- 
especially at short times, which are the focus of our analysis considering the 
projective equivalence detailed in the previous section. 
In particular, for both FM and AFM models, $\Delta S(t)$ starts bending 
downwards at short times as visible in Fig.~\ref{f.FITS_tVMC}.
This means that the first oscillation of $\Delta S$ is in phase with that of $S_{\rm LSW}$ when 
$A_{\bm k} > 0$ (since in that case $A_{\bm k} \cos(2\omega_{\bm k} t)$ bends downwards at short times as well); 
instead it is in phase opposition with that of $S_{\rm LSW}$ when $A_{\bm k}<0$. 
As a consequence, one expects the frequency of the first oscillation of $S(\bm k, t)$ to  increase 
with respect to that of $S_{\rm LSW}(\bm k, t)$ when $A_{\bm k} > 0$, and to  decrease when $A_{\bm k} < 0$. 
This is illustrated in Fig.~\ref{f.FITS_tVMC} when comparing $S_{\rm LSW}(\bm k, t)$ 
to the tVMC and LSW results with OBC (which do contain the $\Delta S$ correction). 
For instance, in the case of the FM, for $(n_x,n_y) = (3,1)$ the first dip of $S(\bm k, t)$ 
occurs before that of  $S_{\rm LSW}(\bm k, t)$, while for $(n_x,n_y) = (4,1)$ 
the first peak of $S(\bm k, t)$ occurs after that of  $S_{\rm LSW}(\bm k, t)$. 

In the AFM case, one observes a  shift of the first peak of $S(\bm k, t)$ to later times compared to that of 
$S_{\rm LSW}(\bm k, t)$, \emph{e.g.} for $(n_x,n_y) = (4,3)$. Yet we also observe a first dip of $S(\bm k, t)$ 
appearing systematically earlier than that of $S_{\rm LSW}(\bm k, t)$ for several wavevectors close to the BZ center. 
This shift can only be partly accounted for by the $\Delta S$ correction, and it rather comes from 
effects beyond linear spin-wave theory. This aspect is also witnessed by the systematic disagreement 
between tVMC data and (corrected) LSW data  for OBCs.

\subsubsection{Influence on the extracted frequencies} 

The $\Delta S$ correction, when comparable to the amplitude of the oscillations of $S(\bm k, t)$, 
has a thus tangible effect on the frequencies $\omega_{\bm k}$ extracted from a short-time fit 
of the tVMC or experimental data. 
This is most clearly seen in the case of the FM, as shown in Fig.~\ref{f.fit_results}C. 
There, compared to the LSW frequency with mPBCs, the fitted frequency is shifted upwards for $A_{\bm k}>0$, 
while it is systematically shifted downwards for $A_{\bm k}<0$. 
For the FM, we attribute the most significant deviations of the fitted frequencies from the LSW prediction 
to this local constraint stemming from the bounded nature of the spin-1/2 spectrum 
(to be contrasted to the unbounded spectrum of free bosons, at the core of standard LSW theory). 

The effect of the $\Delta S$ correction is less clear on the AF, for which an upward shift 
of the frequencies for $A_{\bm k} > 0$ is observed beyond what can be explained with the correction. 
Yet the upward shift in frequencies is most evident when $|\Delta S| \gtrsim |A_{\bm k}|$. 
On the other hand, the downward shift in frequencies when  $A_{\bm k} < 0$ is far less visible, 
since the negative values of $A_{\bm k}$ can be significantly larger in magnitude than $|\Delta S|$, 
as shown in Fig.~\ref{f.fit_results}B.

\subsubsection{Comparison between different estimates for the frequency}

We conclude this section by comparing the fitted amplitude 
$A_{\bm k}$ and offset $C_{\bm k}$ from tVMC data with the predictions from standard LSW theory. 
In the FM case, the fitted amplitude $A_{\bm k}$ and offset $C_{\bm k}$ agree rather well with the LSW prediction 
-- much better than the fitted frequency $\omega_{\bm k}$. 
Figure~\ref{f.fit_results}A compares the LSW predictions and tVMC fits for the $A_{\bm k}$ amplitudes.
Figure~\ref{f.fit_results}C compares the LSW prediction for the $\omega_{\bm k}$  
extracted  from the fitted  $A_{\bm k}$ and $C_{\bm k}$ coefficients 
using $ J \gamma_0 \sqrt{2C_{\bm k}-1}$
and $J \gamma_0 \sqrt{1-2A_{\bm k}}$.  
In the case of the offset $C_{\bm k}$, the agreement with standard LSW theory is not surprising, 
since we argued in Sec.~\ref{s.corr} that $\Delta S$ averages to a small, ${\cal O}(1/N)$ value. 
This analysis validates the robustness of the frequency extraction from the offset $C_{\bm k}$ whenever 
LSW applies to the description of the dynamics. This is far less obvious for the amplitude $A_{\bm k}$, 
which a priori should be affected by the correction. Indeed one sees a systematic 
(albeit small) shift when $A_{\bm k}$ from LSW theory crosses zero. 
For the AFM, one can see the limit of the LSW prediction for the 
amplitude $A_{\bm k}$ in Fig.~\ref{f.fit_results}B: this comes from physics beyond 
LSW theory which is not accounted for by the $\Delta S$ correction, as already argued above. 
The fitted offset $C_{\bm k}$ maintains instead a rather close correspondence with that of LSW theory, 
as shown in Fig.~\ref{f.fit_results}D.

\section{Density of excitations triggered by the quench, and relationship to the thermodynamics}

\begin{figure*}
\centering
\includegraphics[width=0.8\textwidth]{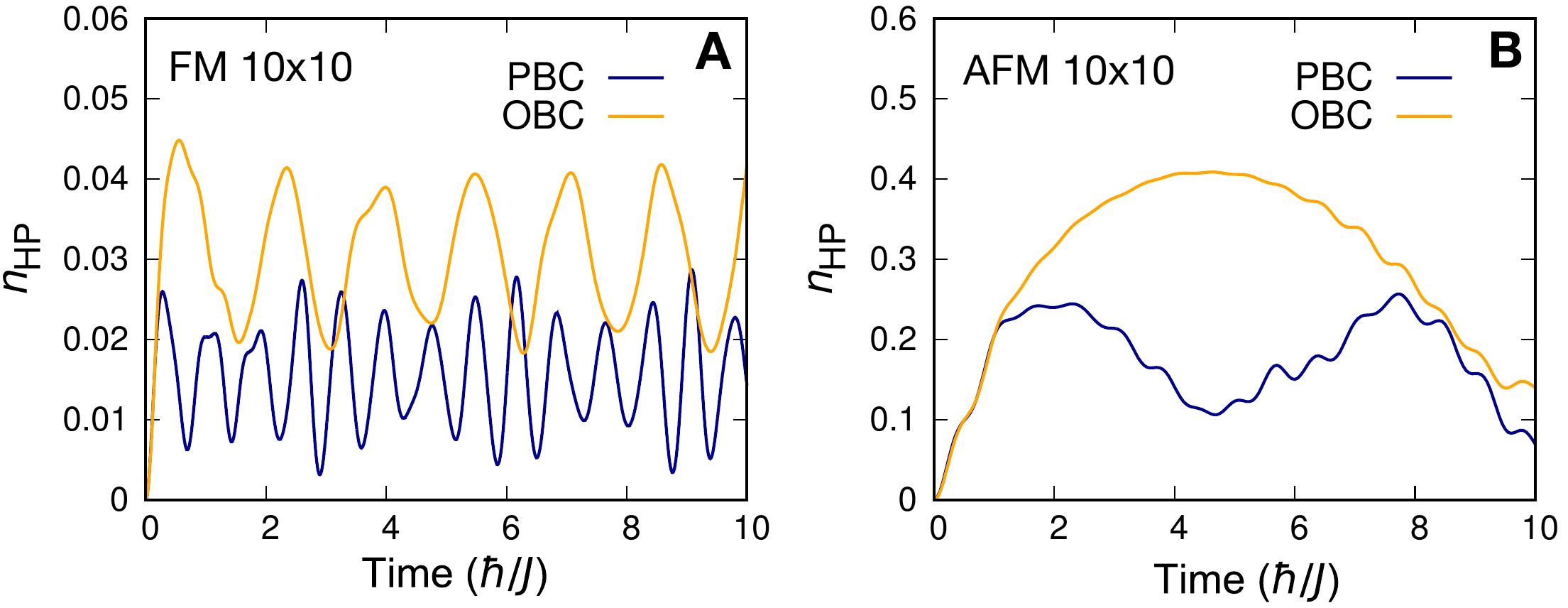}
\caption{{\bf Density of finite-momentum Holstein-Primakoff (HP) bosons generated by the dynamics.}
\textbf{(A)}: time-dependent density of finite-momentum (FM) 
HP bosons generated during the linear spin-wave theory dynamics of the FM on a $10\times 10$ lattice with PBC and OBC;
\textbf{(B)}: same result for the AFM.}
\label{f.nFM}
\end{figure*}

The amount of excitations triggered by the initial quench can be estimated 
via linear spin-wave theory in two alternative ways. 
In the case of periodic boundary conditions, one can simply calculate the density of 
finite-momentum magnons which are present in the initial state, 
$n_m = \frac{1}{N} \sum_{\bm k \neq 0} \langle a_{\bm k}^\dagger a_{\bm k} \rangle= \frac{1}{N} \sum_{\bm k \neq 0} v_{\bm k}^2$. 
For a $10\times 10$ lattice we find that
$n_m  \approx 7.4 \times 10^{-3}$ for the FM  and $n_m \approx  6.6 \times 10^{-2}$ for the AFM. 
The two densities differ by an order of magnitude, exhibiting the enhanced role of fluctuations 
in the case of the AFM compared with the FM. 
The small values of the densities may suggest that neglecting non-linear bosonic terms in the Hamiltonian, 
leading to spin-wave theory, is well justified in both cases.
 
Nonetheless the above observation needs to be reconsidered when calculating instead 
the density of finite-momentum HP bosons generated {\it during} the quench dynamics. 
This density is defined as $n_{\rm HP} = \frac{1}{N} \sum_{\bm k \neq 0} \langle b_{\bm k}^\dagger b_{\bm k} \rangle$; 
the approximation leading to Eq.~\eqref{e.H2real} can be considered to be valid if $n_{\rm HP}$ remains 
much smaller than its maximum value, $2s = 1$,  at least for finite-momentum bosons. 
The zero-momentum ones, corresponding to the zero mode of the U(1) symmetric Hamiltonian, 
can be treated separately as a non-linear rotor variable:
They do not contribute to the correlations in the $S^z$ spin components \cite{Roscilde2023}. 
On the contrary, a large density of finite-momentum HP bosons implies that the linearization 
of the Hamiltonian dynamics is a poor approximation:  
boson-boson interactions must be accounted for, as well as the coupling between 
the zero-momentum bosons and the finite-momentum ones.

The time evolution of $n_{\rm HP}$ is shown in Fig.~\ref{f.nFM}. 
We observe that the FM dynamics develops a density of finite momentum bosons 
reaching peak values of $n_{\rm FM} \approx 0.04$ (with OBC). 
Contrarily, the largest value for the AFM is an order of magnitude higher, bringing the system to a regime in which 
the population of HP bosons can no longer be considered as dilute.  
This shows that non-linearities are expected to play a very different role for the AFM with respect to the FM.

\section{Numerical methods}

\begin{figure*}
\centering
\includegraphics[width=0.8\textwidth]{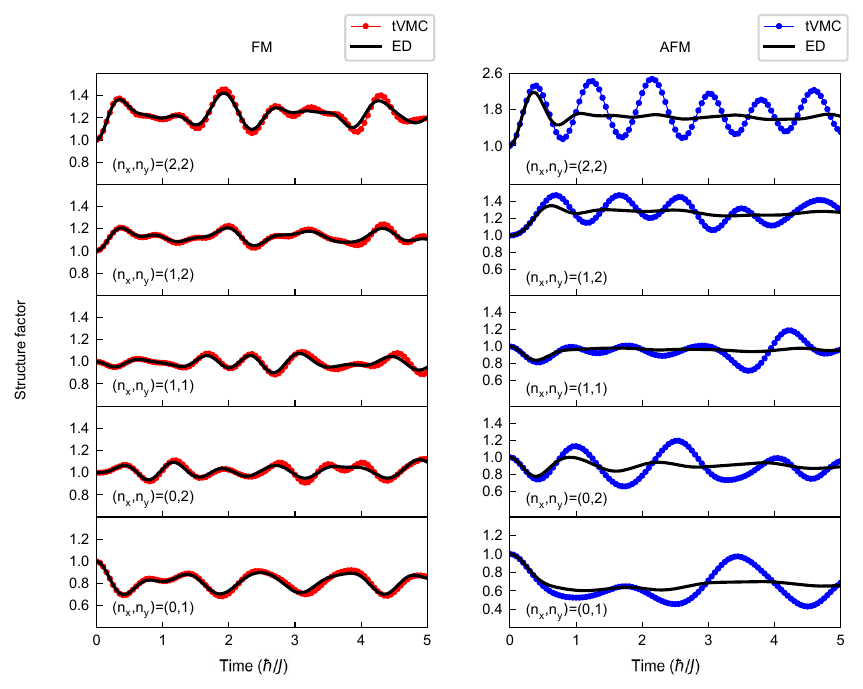}
\caption{{\bf Comparison between tVMC and exact diagonalization}
for the time-dependent structure factor of the dipolar XY model on a $4\times 4$ lattice (with OBC)
for all the non-equivalent wavevectors in the Brillouin zone. 
Left panels: ferromagnet; right panels: antiferromagnet.}
\label{f.tVMC_vs_ED}
\end{figure*}

The theoretical results for the time evolution of dipolar spin Hamiltonians have been obtained using 
time-dependent linear spin-wave theory (discussed in Sec.~\ref{s.LSW}), 
and two numerical approches:
1) exact diagonalization on small ($4 \times 4$) lattices (using the QuSpin package \cite{QuSpin}); 
and 2) time-dependent variational Monte Carlo (tVMC). 
The latter approach is based on the time evolution of a trial wavefunction 
governed by the time-dependent variational principle \cite{bookTDVP}. 
We used a spin-Jastrow (or pair-product) wavefunction \cite{Thibautetal2019,Comparin2022} 
for the present study, as this wavefunction has already proven to be very successful in reproducing 
the non-equilibrium physics of the dipolar XY ferromagnet \cite{Comparin2022, Bornet2023} 
(tested by comparison with exact diagonalization on small systems): 
Fig.~\ref{f.tVMC_vs_ED}A shows the tVMC predictions for $S(\bm k,t)$ 
of the FM with OBC together with exact diagonalization ($4 \times 4$
lattice), confirming its ability to grasp the correct behavior 
over the entire time scale relevant for the experiment. 
As a consequence, we use tVMC to extend the numerical predictions 
beyond exact diagonalization, and up to the $10\times 10$ OBC lattices implemented in the experiment. 

In the case of the dipolar antiferromagnet, though, the ability of the Jastrow wavefuction 
to reproduce the correct non-equilibrium evolution is much more limited. 
When comparing with exact diagonalization (see Fig.~\ref{f.tVMC_vs_ED}B), 
tVMC based on Jastrow wavefunction reproduces correctly the early-time dynamics 
-- most importantly for us, the first oscillations of the time-dependent structure factor -- 
until a time of order $\hbar/J$. However, the Jastrow wavefunction misses the later decay of the oscillations. 
Hence we can use the tVMC results to extract characteristic wavevector-dependent 
frequencies in the early-time dynamics, independently of spin-wave theory. 

In the case of the ferromagnet, we can also make quantitative predictions 
for the thermal state towards which the dynamics is expected to relax at long times. 
This state should be dictated uniquely by the conserved quantities in the dynamics: 
If the energy is the only conserved quantity, the thermal state is given by the Gibbs ensemble (GE) 
at the temperature $T_{\rm CSS}$ such that  
$\langle {\rm CSS} | H | {\rm CSS} \rangle = \langle H \rangle_{T_{\rm CSS}}$, 
where $\langle ... \rangle_T = {\rm Tr}[e^{-H/(k_BT)} (...)]/{\rm Tr}[e^{-H/(k_BT)}]$ 
is the GE average at temperature $T$. 
The dynamics of the dipolar XY model also conserves the collective spin 
$J^z = \sum_i \sigma_i^z$ and all its functions. 
Of particular interest for us is $\langle {\rm CSS} | (J^z)^2 | {\rm CSS} \rangle = N$,
which corresponds to the structure factor at zero wavevector $S(0) = \langle (J^z)^2 \rangle/N$. 
Hence, in order to correctly capture the structure factor of the equilibrium state to 
which the dynamics should relax, we run quantum Monte Carlo (QMC) simulations 
(based on the stochastic series expansion approach \cite{SyljuasenS2002}) in the generalized Gibbs 
ensemble (GGE), in which averages are calculated as 
$\langle ... \rangle_{T,\lambda} = {\rm Tr}\{e^{-[H/(k_BT) - \lambda (J^z)^2]} (...) \} /
{\rm Tr}\{e^{-[H/(k_BT) - \lambda (J^z)^2]} \}$, 
where $\lambda$ is a Lagrange multiplier tuned so that  $\langle (J^z)^2 \rangle_{T,\lambda} = N$.  
Our GGE QMC simulations are performed on a $N=100$ array with 
open boundary conditions, in order to mimic closely the experiment. 
For this lattice size, the temperature matching the initial-state energy is $T/J \approx 1.2$.

Some finite-temperature calculations are possible for the dipolar antiferromagnet 
using tensor network methods, albeit at restricted system sizes.
Following the methodology described in Ref.~\cite{Chen2023}, we use minimally 
entangled typical thermal states to estimate an energy-temperature calibration. 
For a $4\sqrt{2}\times 8\sqrt{2} = 64$ site cylinder, we find $T_{\mathrm{CSS}}^{\mathrm{AFM}}/J \approx 0.6$. 
As a benchmark, carrying out the same procedure for the nearest-neighbor model gives a value of 
$T_{\mathrm{CSS}}^{\mathrm{nearest-neighbor}}$ differing by about 10$\%$ from prior QMC 
results~\cite{ding1992phase}.
We attribute this discrepancy to strong finite-size effects near the Berezinskii-Kosterlitz-Thouless 
transition -- namely, our system size is comparable to the correlation length $\xi/a\approx 23$ 
found at $T_{\mathrm{CSS}}^{\mathrm{nearest-neighbor}}$~\cite{ding1992phase}.

\section{Ground-state spectral functions vs. quench spectroscopy for dipolar XY models} 

We discuss here the spectral function accessible to quench spectroscopy, 
and compare it with more conventional spectral functions 
(i.e. the dynamical structure factor and its generalizations), 
obtained by probing the linear response of the system at thermal equilibrium. 
In particular, using exact diagonalization to calculate the conventional spectral functions, 
we gather informations on the nature of the spectrum of elementary excitations in the dipolar XY models. 

\subsection{Quench spectral function vs. dynamical structure factors} 

The Fourier transform of the time-dependent structure factor given by Eq.~\eqref{e.Skt}, 
monitored over an infinite time, reconstructs the quench spectral function: 
\begin{equation}
{\cal Q}(\bm k, \omega) = 
\sum_{nm}  \langle \Psi(0)| n \rangle \langle m | \Psi(0)\rangle \langle n | \sigma_{\bm k}^z \sigma_{-\bm k}^z | m\rangle 
\delta(\omega-\omega_{nm})~.
\end{equation}
As discussed in Sec.~\ref{s.QS}, this spectral function contains contributions from all the transitions 
$|m\rangle \to |n\rangle$ among states which are connected by two spin flips at opposite momenta 
and which have a significant overlap with the initial state of the quench dynamics. 

Contrarily, the conventional spectral function associated with $\sigma^z_{\bm k}$ operator, 
and relevant for spectroscopy at thermal equilibrium, is the {\it one-spin-flip dynamical structure factor}: 
\begin{equation}
{\cal S}_1(\bm k,\omega) = \frac{1}{\cal Z} \sum_{nm} e^{-\beta E_m} |\langle n | \sigma_{\bm k}^z |m \rangle|^2 
\delta (\omega-\omega_{nm})
\end{equation}
where $\beta = (k_B T)^{-1}$ is the inverse temperature. This structure factor is probed e.g. by neutron scattering, 
as it dictates the scattering cross sections of neutrons from magnetic materials at thermal equilibrium \cite{Lovesey1984}. 
At variance with the quench spectral function ${\cal Q}(\bm k, \omega)$, 
${\cal S}_1(\bm k,\omega)$ probes transitions among states connected by
a {\it single} spin-flip operator $\sigma^z_{\bm k}$. Here, the final state of the transitions needs not be thermally populated. 

It is then instructive to consider a {\it two-spin-flip dynamical structure factor}, defined as 
\begin{equation}
{\cal S}_2(\bm k,\omega) = \frac{1}{\cal Z} \sum_{nm} e^{-\beta E_m} |\langle n | \sigma_{\bm k}^z \sigma_{-\bm k}^z  |m \rangle|^2 \delta (\omega-\omega_{nm})
\end{equation}
which probes instead the same  transitions as those that contribute to the quench 
spectral function, although not with the same weights. 
Related spectral functions are measured in condensed-matter experiments by two-magnon 
Raman scattering \cite{Deveraux2007} and resonant inelastix X-ray scattering \cite{Ament2011}. 

In the following we contrast these three spectral functions in the case of the dipolar XY models. 

\subsection{Spectral functions for dipolar XY magnets from exact diagonalization} \label{s.spectral}

\begin{figure*}
\centering
\includegraphics[width=0.8\textwidth]{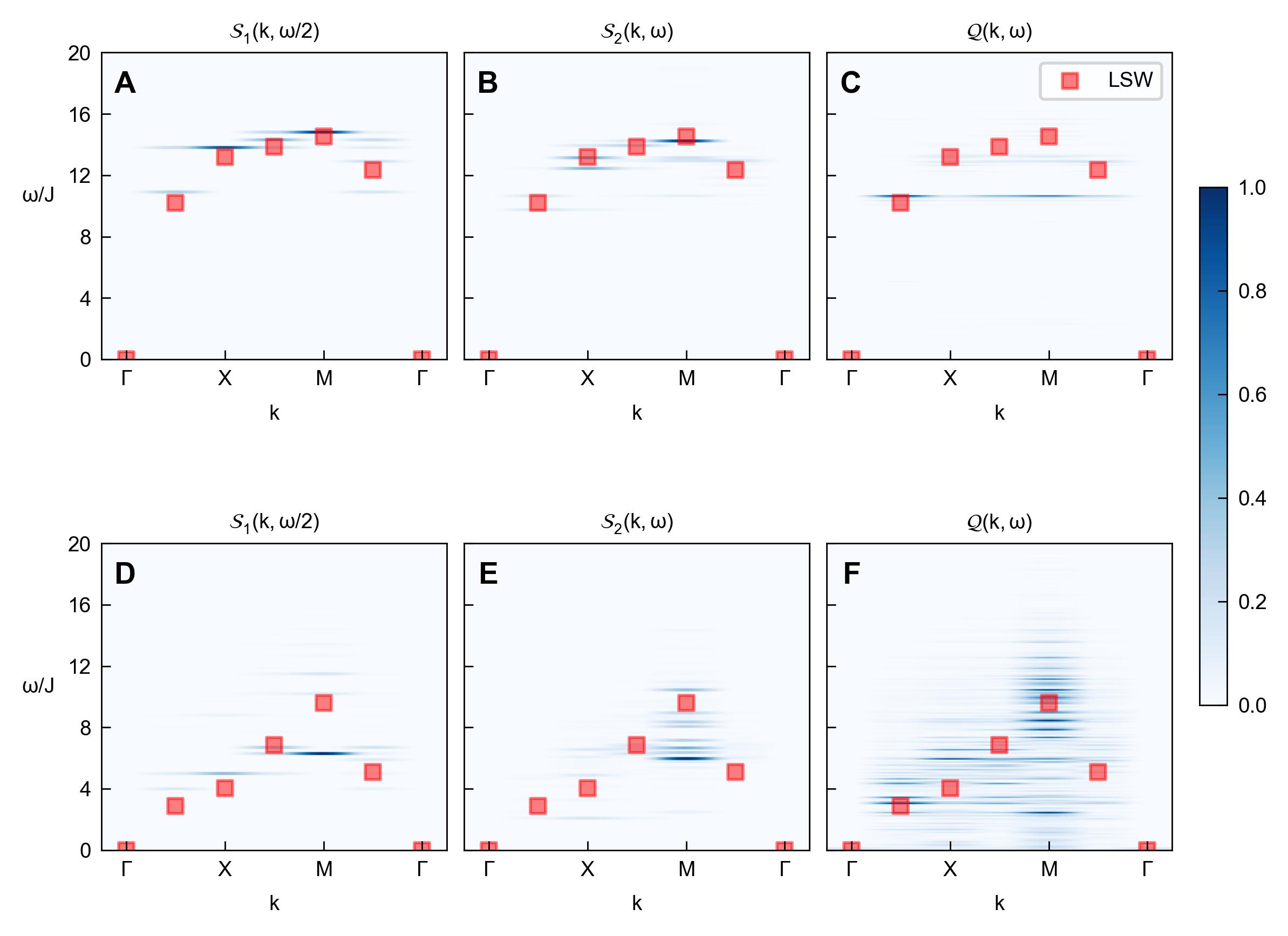}
\caption{{\bf Spectral functions} for the $4 \times 4$ dipolar XY model (with PBC). 
\textbf{(A)},\textbf{(D)}: Zero-temperature one-spin-flip dynamical structure factor ${\cal S}_1(\bm k,\omega)$ for the 
FM (A) and the AFM (D); 
\textbf{(B)},\textbf{(E)} zero-temperature two-spin-flip dynamical structure factor ${\cal S}_2(\bm k,\omega)$ for the 
FM (B) and the AFM (E); 
\textbf{(C)},\textbf{(F)} Quench spectral function (referred to the coherent spin state) ${\cal Q}(\bm k, \omega)$ for the 
FM (C) and the AFM (F). 
The ${\bm k}$ vectors are taken along the the path indicated in Fig.~4 
of the main text. The red squares mark the dispersion relation from linear spin-wave theory. 
In all the panels, the values of the spectral functions are normalized to their maximum.}
\label{f.spectral}
\end{figure*}

Fig.~\ref{f.spectral} shows the one- and two-spin-flip dynamical structure factors at $T=0$, 
as well as the quench spectral function, for both the ferromagnetic and antiferromagnetic dipolar XY models 
on a $4\times 4$ lattice with PBC, obtained via exact diagonalization. 
The $T=0$ dynamical structure factors reveal the spectrum of excitations created 
by the spin-flip operators $\sigma_{\bm k}^z$ and $\sigma_{\bm k}^z \sigma_{-\bm k}^z$ 
onto the Hamiltonian ground state. 

Fig.~\ref{f.spectral}A,B show that, for the dipolar ferromagnet, 
the spectrum of two-spin-flip excitations corresponds to twice the spectrum for the single-spin-flip ones 
-- highlighting the free quasiparticle nature of the excitations themselves. 
This is further confirmed by the correspondence between the characteristic frequencies of the one-spin-flip 
and two-spin-flip spectra compared to the predictions of linear spin-wave theory. 
The quench spectral function (Fig.~\ref{f.spectral}C) has also a similar structure 
to that of the two-spin-flip dynamical structure factor, but with additional lower frequencies: 
These correspond to transitions which do not involve the Hamiltonian ground state. 

A different picture emerges for the dipolar antiferromagnet, as shown in Figs.~\ref{f.spectral}D,F. 
The one-spin-flip structure factor (Fig.~\ref{f.spectral}D) exhibits multiple frequencies for the same 
wavevector with similar spectral weight. For the wavevector $(\pi,\pi)$ ($M$-point) a dominant frequency 
emerges, which nonetheless differs significantly from the spin-wave prediction. 
As we discuss in Sec.~\ref{s.decay}, the first nonlinear correction to spin-wave theory 
leads to a finite decay rate of single-magnon excitations at $(\pi,\pi)$, 
which may indeed be reflected in the result on the small size accessible to exact diagonalization.  
This already suggests a significant departure of the elementary excitation spectrum 
from that of free quasiparticles. Yet the most dramatic departure is seen at the level of 
the two-spin-flip structure factor (Fig.~\ref{f.spectral}E), which shows an even broader spectrum 
of frequencies at essentially all wavevectors, and especially so at $(\pi,\pi)$, 
where  a ``continuum" of frequencies emerges. 
This result suggests that two-magnon-decay processes may be even more prominent that single-magnon ones. 
The broad spectral features observed in the two-spin-flip structure factor are much more 
prominently revealed in the quench spectral function (Fig.~\ref{f.spectral}F), 
whose structure is consistent with the fast decay of oscillations of the time-dependent 
structure factor revealed by the experiment, as discussed in the main text. 

\section{One-magnon decay}\label{s.decay}

\begin{figure*}
\centering
\includegraphics[width=0.8\textwidth]{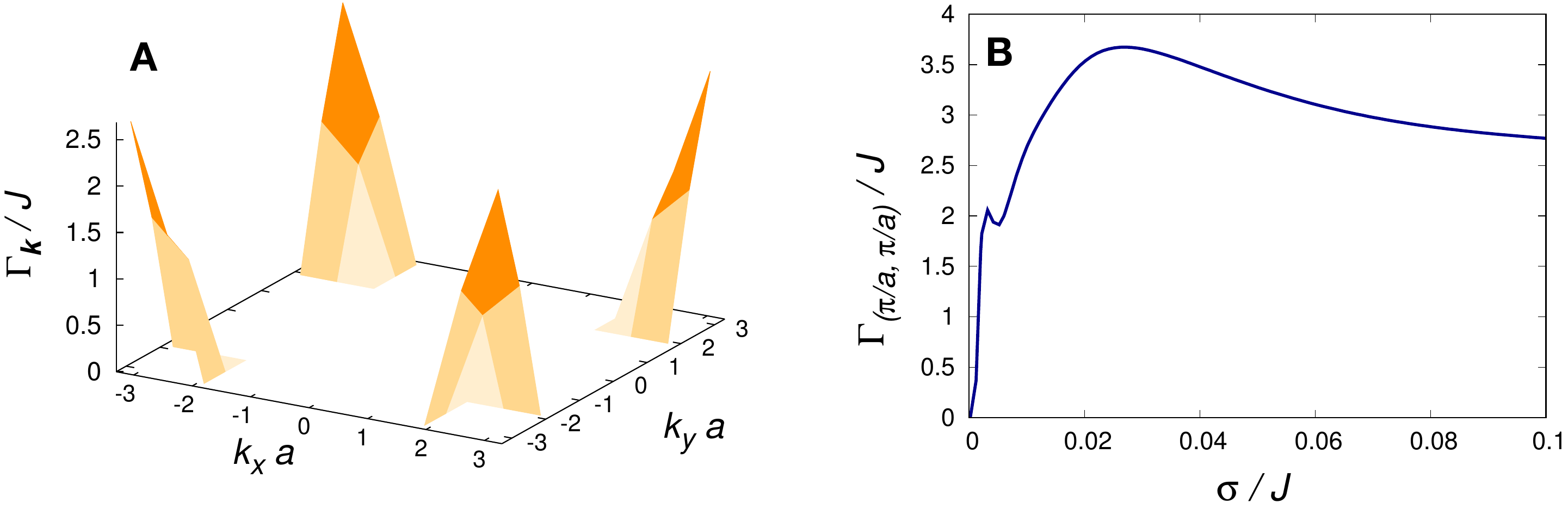}
\caption{{\bf Single-magnon decay rate}. 
\textbf{(A)} $\Gamma_{\bm k}$ for the spin-wave modes of a $10 \times 10$ dipolar XY antiferromagnet. 
The Dirac function entering in Eq.~\eqref{e.Gammak} has been broadened 
to a Gaussian with width $\sigma = 0.01 J$. 
\textbf{(B)} $\sigma$-dependence of the decay rate for the spin-wave mode at ${\bm k} = (\pi/a,\pi/a)$.}
\label{f.Gammak}
\end{figure*}

To test the robustness of spin-wave predictions beyond the harmonic approximation 
we consider the first nonlinear correction to the quadratic Hamiltonian Eq.~\eqref{e.H2real}, 
and calculate its effect on possible decay of single-magnon 
excitations using Fermi's golden rule \cite{ChernyshevZ2013}. 
The first  nonlinear correction to the quadratic Hamiltonian is the quartic Hamiltonian, 
expressed in momentum space:
\begin{eqnarray}
H_4 =  \frac{1}{N} \sum_{\bm k \bm k' \bm q}  
\Big [ &{\cal A}_{\bm k \bm k' \bm q} & b_{\bm k-\bm q}^\dagger  b_{\bm k'+\bm q}^\dagger b_{\bm k} b_{\bm k'} \nonumber \\
+& {\cal B}_{\bm k  \bm k' \bm q} & 
\Big ( b_{\bm k}^\dagger b_{\bm k'}^\dagger b_{\bm q - \bm k - \bm k'}^\dagger b_{\bm q} + {\rm h.c.} \Big ) ~\Big] 
\label{e.H4}
\end{eqnarray}
where 
${\cal A}_{\bm k \bm k' \bm q}  =  J_{\bm q} - (J_{\bm k'} + J_{\bm k - \bm q})/4$
and ${\cal B}_{\bm k \bm k' \bm q}  =  -(J_{\bm k} + J_{\bm q - \bm k - \bm k'})/8$. 
It can be Bogolyubov transformed to a quartic Hamiltonian in terms of the magnon operators 
$a_{\bm k}, a_{\bm k}^\dagger$. The resulting expression is rather lengthy but straightforward to obtain. 
In a system with periodic boundary conditions, this Hamiltonian leads to a decay process 
of a single magnon into three magnons via terms of the form 
$a^\dagger_{{\bm q}_1} a^\dagger_{{\bm q}_2} a^\dagger_{{\bm q}_3} a_{\bm k}$, 
conserving momentum (${\bm k} = {\bm q}_1 + {\bm q}_2 + {\bm q}_3$) and 
energy ($\omega_{\bm k} = \omega_1 + \omega_2 + \omega_3$), with a rate  \cite{ChernyshevZ2013}:
\begin{align} 
\Gamma_{\bm k} =  
\frac{\pi}{8\hbar} \sum_{\bm q_1, \bm q_2}  
\langle 1_{\bm q_1}, 1_{\bm q_2}, 1_{\bm k - \bm q_1 - \bm q_2} | H_4 | 1_{\bm k} \rangle|^2 \label{e.Gammak} \\
\delta(\omega_{\bm k} - \omega_1 - \omega_2 - \omega_3)    \nonumber
\end{align}
where the states $|\{n_{\bm q} \}\rangle$ are Fock states of magnon occupation. 
To calculate the decay rate on a finite-size system, the $\delta$ function of Eq.~\eqref{e.Gammak} 
is given a finite width, e.g. by approximating it with a Gaussian 
$\delta(x) \to  \exp[-x^2/(2\sigma^2)]/(\sqrt{2\pi}\sigma)$.  
The finite (i.e. nonzero) width corresponds to a finite (i.e. not infinite) perturbation time, 
which, within time-dependent perturbation theory, leads to an uncertainty in the zero frequency of the perturbation.
Finite observation times do not break energy conservation but introduce extra Fourier components that lead to broadening.
The numerical value of $\Gamma_{\bm k}$ depends on $\sigma$, but only weakly when $\sigma$ 
is sufficiently large (\emph{e.g.} $\sigma/J \gtrsim 5\times 10^{-3}$ 
for a $10 \times 10$ system, as shown in Fig.~\ref{f.Gammak}B).

In the case of the dipolar ferromagnet, the calculation of the decay rate leads to a negligible result -- 
consistent with the robustness of linear spin-wave theory.  
Contrarily, in the antiferromagnet case, a significant decay rate emerges for ${\bm k} \approx (\pi,\pi)$, 
as shown in Fig.~\ref{f.Gammak}A. The propensity to decay of antiferromagnetic magnons can be 
understood by inspecting the dispersion relation of Fig.~\ref{f.LSW_dispersion}: 
such a dispersion relation  is nearly linear close to the diagonal of the Brillouin 
zone connecting the origin to the $(\pi,\pi)$ corner. This allows for the
decay channels ${\bm k} \to  {\bm q}_1 + {\bm q}_2 + {\bm q}_3$ which conserve energy, 
and which are more numerous the closer ${\bm k}$ is to $(\pi,\pi)$. 

The latter result suggests that the decay of oscillations in the time-dependent structure 
factor observed in the experiment can be -- at least partially --  traced back to the decay 
process of single-magnon excitations. 
Nonetheless the initial state of the experiment -- the coherent spin state along $y$ -- 
corresponds to a finite density of magnons, so that multi-magnon decay processes are 
expected to be as relevant as single-magnon ones. In fact, the two-spin-flip dynamical 
structure factor shown in Sec.~\ref{s.spectral} shows a richer frequency content than the 
single-spin-flip one, which suggests already the importance of two-magnon decay processes 
compared to single-magnon ones. A full study of the (multi-)magnon decay processes 
quantitatively explaining the damping of oscillations observed in the experiment 
goes far beyond the scope of the present work. 
Yet this aspect deserves further investigations, in order to understand more deeply the 
insight into the nature of elementary excitations which is provided by quench spectroscopy.

\end{document}